\documentclass[%
reprint,
superscriptaddress,
amsmath,
amssymb,
aps,
floatfix,
citeautoscript,
longbibliography
]{revtex4-1}

\usepackage{graphicx}
\usepackage{dcolumn}
\usepackage{bm}
\usepackage{bbm}
\usepackage{units}
\usepackage{braket}
\usepackage{epstopdf}
\usepackage{color}
\usepackage{float}
\usepackage{csquotes}
\usepackage{array}

\usepackage[colorlinks, linkcolor=blue, citecolor=blue, urlcolor=blue, breaklinks=true]{hyperref}

\newcommand{\kb}[1]{|#1\rangle\langle #1|}
\newcommand{\kbsub}[2]{|#1\rangle_{#2}\langle #1|}
\newcommand{\ketab}[2]{|#1_{\mathrm{A}}#2_{\mathrm{B}}\rangle}

\begin{document}

\title{Experimental Single-Copy Entanglement Distillation}

\author{Sebastian Ecker}
\email{sebastian.ecker@oeaw.ac.at}
\affiliation{Institute for Quantum Optics and Quantum Information (IQOQI), Austrian Academy of Sciences, Boltzmanngasse 3, 1090 Vienna, Austria}
\affiliation{Vienna Center for Quantum Science and Technology (VCQ), Faculty of Physics, University of Vienna, Boltzmanngasse 5, 1090 Vienna, Austria}

\author{Philipp Sohr}
\affiliation{Institute for Quantum Optics and Quantum Information (IQOQI), Austrian Academy of Sciences, Boltzmanngasse 3, 1090 Vienna, Austria}
\affiliation{Vienna Center for Quantum Science and Technology (VCQ), Faculty of Physics, University of Vienna, Boltzmanngasse 5, 1090 Vienna, Austria}

\author{Lukas Bulla}
\affiliation{Institute for Quantum Optics and Quantum Information (IQOQI), Austrian Academy of Sciences, Boltzmanngasse 3, 1090 Vienna, Austria}
\affiliation{Vienna Center for Quantum Science and Technology (VCQ), Faculty of Physics, University of Vienna, Boltzmanngasse 5, 1090 Vienna, Austria}

\author{\\Marcus Huber}
\affiliation{Institute for Quantum Optics and Quantum Information (IQOQI), Austrian Academy of Sciences, Boltzmanngasse 3, 1090 Vienna, Austria}
\affiliation{Institute for Atomic and Subatomic Physics, Vienna University of Technology, 1020 Vienna, Austria}

\author{Martin Bohmann}
\affiliation{Institute for Quantum Optics and Quantum Information (IQOQI), Austrian Academy of Sciences, Boltzmanngasse 3, 1090 Vienna, Austria}
\affiliation{Vienna Center for Quantum Science and Technology (VCQ), Faculty of Physics, University of Vienna, Boltzmanngasse 5, 1090 Vienna, Austria}

\author{Rupert Ursin}
\email{rupert.ursin@oeaw.ac.at}
\affiliation{Institute for Quantum Optics and Quantum Information (IQOQI), Austrian Academy of Sciences, Boltzmanngasse 3, 1090 Vienna, Austria}
\affiliation{Vienna Center for Quantum Science and Technology (VCQ), Faculty of Physics, University of Vienna, Boltzmanngasse 5, 1090 Vienna, Austria}

\begin{abstract}
The phenomenon of entanglement marks one of the furthest departures from classical physics and is indispensable for quantum information processing. 
Despite its fundamental importance, the distribution of entanglement over long distances through photons is unfortunately hindered by unavoidable decoherence effects. 
Entanglement distillation is a means of restoring the quality of such diluted entanglement by concentrating it into a pair of qubits. Conventionally, this would be done by distributing multiple photon pairs and distilling the entanglement into a single pair. Here, we turn around this paradigm by utilising pairs of single photons entangled in multiple degrees of freedom.
Specifically, we make use of the polarisation and the energy-time domain of photons, both of which are extensively field-tested.
We experimentally chart the domain of distillable states and achieve relative fidelity gains up to \unit[13.8]{\%}. 
Compared to the two-copy scheme, the distillation rate of our single-copy scheme is several orders of magnitude higher, paving the way towards high-capacity and noise-resilient quantum networks.
\end{abstract}

\maketitle

    Entanglement lies at the heart of quantum physics, reflecting the quantum superposition principle between remote subsystems without a classical counterpart.
    In addition to its fundamental importance, entanglement is an essential resource for most quantum information applications \cite{wilde2013quantum}, and its distribution between remote parties provides the basis for quantum communication~\cite{xu20}, distributed quantum computing~\cite{cuomo20}, and eventually the quantum internet~\cite{wehner18}.
    For entanglement distribution, photons are almost ideal carriers of quantum states.
    It is possible to create close-to-maximally entangled photon-pairs at high rates in the laboratory~\cite{steinlechner12,chen18,ecker2021,kues2017}, and recent efforts are venturing out of laboratory environments \cite{yin17,joshi20,valivarthi16,Wengerowsky19}, bringing the ultimate goal of a global quantum network within reach.
    Nonetheless, noise and interaction with the environment are unavoidable, leading to decoherence \cite{schlosshauer19} and the degradation of entanglement. 
    
\begin{figure}[b]
        \centering
        \includegraphics[width=\columnwidth]{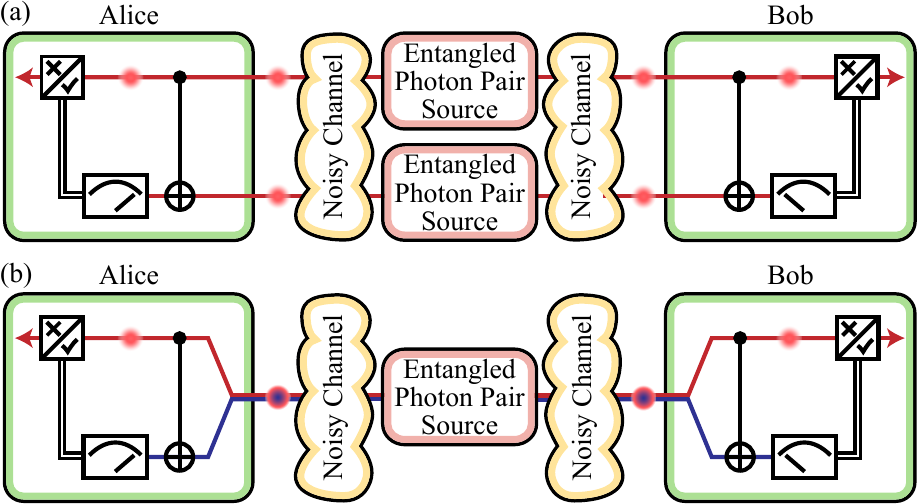}
        \caption{Schematics of an elementary entanglement distillation step.
        (a) Two-copy entanglement distillation in which both Alice and Bob apply a CNOT gate between the two single photons they receive.
        The control photon pair is successfully purified if the measurement outcomes of the target qubits are correlated.
        (b) Single-copy entanglement distillation employs two entangled subspaces encoded in DOF of a single photon pair.
        The CNOT gate is now applied between the two DOF and the successful distillation of the control (upper) DOF is heralded by the measurement outcomes of the target (lower) DOF.
        The classical channel is omitted.}
        \label{fig:scheme}
    \end{figure}

    To counteract such detrimental noise effects, entanglement distillation, also referred to as entanglement purification, was introduced~\cite{bennett96,deutsch96}.
    In entanglement distillation, two copies of a noisy entangled state are employed to distill a single copy with a higher degree of entanglement, a process which can be cascaded until ultimately a pure Bell state is reached.
    Its implementation is based on two-photon controlled NOT (CNOT) gates [see Fig.~\hyperref[fig:scheme]{\ref*{fig:scheme}(a)}], and facilitates, for example, quantum repeaters ~\cite{briegel98,duer99,chen17}.
    Experimental implementations of entanglement distillation, however, face two challenges:
    firstly, high optical losses in realistic scenarios lead to low transmission probabilities of a single photon pair~\cite{yin17,Wengerowsky19}, let alone the transmission of multiple photon pairs, which substantially limits the distillation rate.
    Secondly, the required two-photon CNOT gate cannot be deterministically realised with passive linear optics~\cite{obrien03,pittman03,gasparoni04,zhao05}.
    Therefore, all demonstrations of photonic two-copy entanglement distillation ~\cite{pan03,yamamoto03,walther05,chen17} are based on a parity check rather than on a genuine CNOT gate.
    \begin{figure*}[t]
    \centering
    \includegraphics[width=1\textwidth]{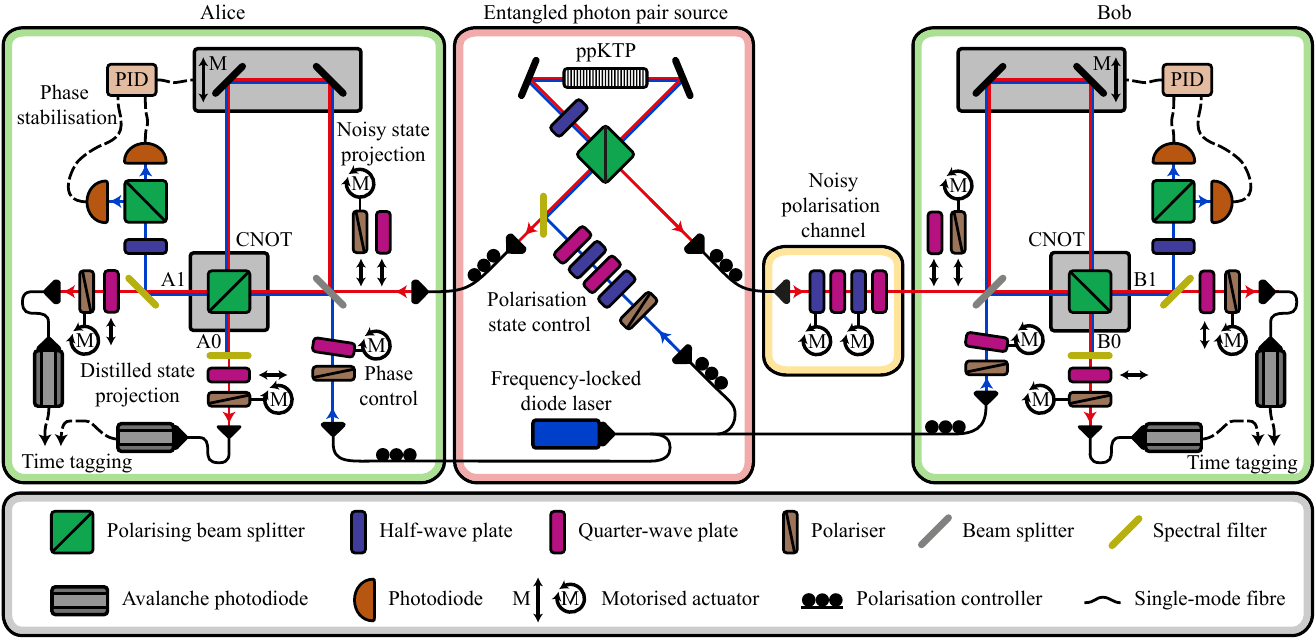}
    \caption{Experimental setup for single-copy entanglement distillation in polarisation and energy-time. Pairs of entangled photons are created in a periodically-poled potassium titanyl phosphate (ppKTP) crystal placed in a Sagnac interferometer. Polarisation entanglement is produced by bidirectionally pumping the crystal in the interferometer and overlapping the resulting photon-pair modes, while a narrow-line width pump laser gives rise to energy-time entanglement. The entangled photons are single-mode-coupled and guided to Alice and Bob, where the distillation is carried out by interfering the polarisation and the energy-time domain of single photons with a modified Franson interferometer, stabilised by a proportional–integral–derivative (PID) controller. The quantum state is either measured before (Noisy state projection) or after (Distilled state projection) the interferometer and the photons collected in paths A0, A1, B0 and B1 are detected and time-tagged.}
    \label{fig:exp_setup}
    \end{figure*}

    To overcome these problems, single-copy entanglement distillation was proposed, which harnesses entanglement in different degrees of freedom (DOF) of a single photon pair~\cite{simon02}, known as hyperentanglement~\cite{Barreiro05,barbieri05,Li16}.
    Instead of operating on two photons, each carrying a qubit, the CNOT gate now acts on two qubits encoded in different DOF of a single photon [see Fig.~\hyperref[fig:scheme]{\ref*{fig:scheme}(b)}].
    Importantly, the CNOT gate between two DOF can be realised deterministically with linear optics \cite{fiorentino04,barreiro08}, which has also been used in a recent purifcation implementation \cite{hu21}.
    Furthermore, hyperentanglement is readily available in spontaneous parametric down-conversion (SPDC) \cite{Kwiat97,Barreiro05} as well as in other photon-pair creating processes \cite{prilm18,reimer19}, and serves as a versatile experimental platform, featuring enhanced communication channel capacity \cite{barreiro08,graham15,williams17}.
    Notably, the polarisation and the energy-time DOF are particularly robust quantum information carriers and have been distributed over free-space \cite{steinlechner17,Jin:19,yin17} and long-distance fibre \cite{marcikic04,Wengerowsky19} links, marking them as ideal candidates for future in-field applications.
    On the other hand, the originally proposed spatial encoding \cite{simon02} is less noise resilient outside of a protected laboratory environment \cite{krenn15,dalio20}.
    
    Here, we report the first experimental implementation of a single-copy distillation protocol exploiting hyperentanglement in the field-tested polarisation and energy-time degrees of freedom.
    We overcome the two principal limitations in standard distillation schemes:
    the probabilistic nature of multi-photon interactions and the low success rates due to two-pair transmission, both obliterating entangled-photon rates in any realistic scenario.
    By introducing different noise scenarios at finely tuned noise levels, we thoroughly test our distillation scheme and successfully recover entanglement and state fidelity.
    We beat the standard two-copy distillation rate by several orders of magnitude and thus unlock quantum communication in unprecedented noise regimes.

\section*{Results}
Our experimental platform consists of an entangled photon-pair source, which is connected to the spatially separated communicating parties Alice and Bob via 12m-long single-mode fibres [Fig.~\ref{fig:exp_setup}].
Alice and Bob each have a distillation setup at their disposal and can characterise their part of the entangled quantum state prior to or after the distillation step.
We obtain polarisation entanglement in our photon-pair source by superposing a SPDC process in the clockwise and the counterclockwise direction of a Sagnac interferometer~\cite{kim06,fedrizzi07}, creating a superposition of an H-polarised and a V-polarised photon pair. 
Energy-time entanglement, on the other hand, arises from energy conservation in the SPDC process driven by a temporally coherent pump field, leading to a potentially large superposition of temporal modes $\ket{t_it_i}$ of the photon pairs \cite{martin17}.
Considering a two-dimensional subspace of the energy-time domain, the resulting hyperentangled state we produce is close to 
\begin{equation}
\ket{\boldsymbol{\Phi}^+_{\mathrm{}}}=
\frac{1}{2}\big[\left(\ket{\text{H,H}}+\ket{\text{V,V}}\right) \otimes \left(\ket{t_\text{L},t_\text{L}}+\ket{t_\text{S},t_\text{S}}\right)\big],
\label{eq:pure_state}
\end{equation}
where both subspaces are entangled in a $\Phi^+$ Bell state.
In order to access the temporal modes at Alice and Bob, we map both photons to the path domain in a Franson interferometer~\cite{franson89} by employing two Mach-Zehnder interferometers with a temporal imbalance of $t_\text{L}-t_\text{S}$ between the long (L) and the short (S) arm.
This mapping is probabilistic due to the randomness of the employed 50:50 beam splitters.
Owing to their indistinguishability, only those photon pairs arriving simultaneously at the outputs of the interferometers exhibit quantum interference, requiring postselection on coincidences~\cite{kwiat93}.
The polarising beam splitter at the output ports of the interferometer constitutes the single-photon CNOT gate.
Depending on the polarisation state (control qubit) of the photon, the path (target qubit) of the photon is either left unchanged or acted upon with an exclusive OR gate.
In analogy with the two-copy distillation protocol \cite{bennett96}, the polarisation state is successfully distilled if the measurement outcomes in the computational path basis are correlated.
For our experimental setup [Fig.~\ref{fig:exp_setup}], this implies that only photons collected in paths A0 and B0 or A1 and B1 are postselected (see Appendix Sec.~\ref{sec:expmethods} for  experimental methods).

\begin{figure}[t]
    \centering
    \includegraphics[width=1\columnwidth]{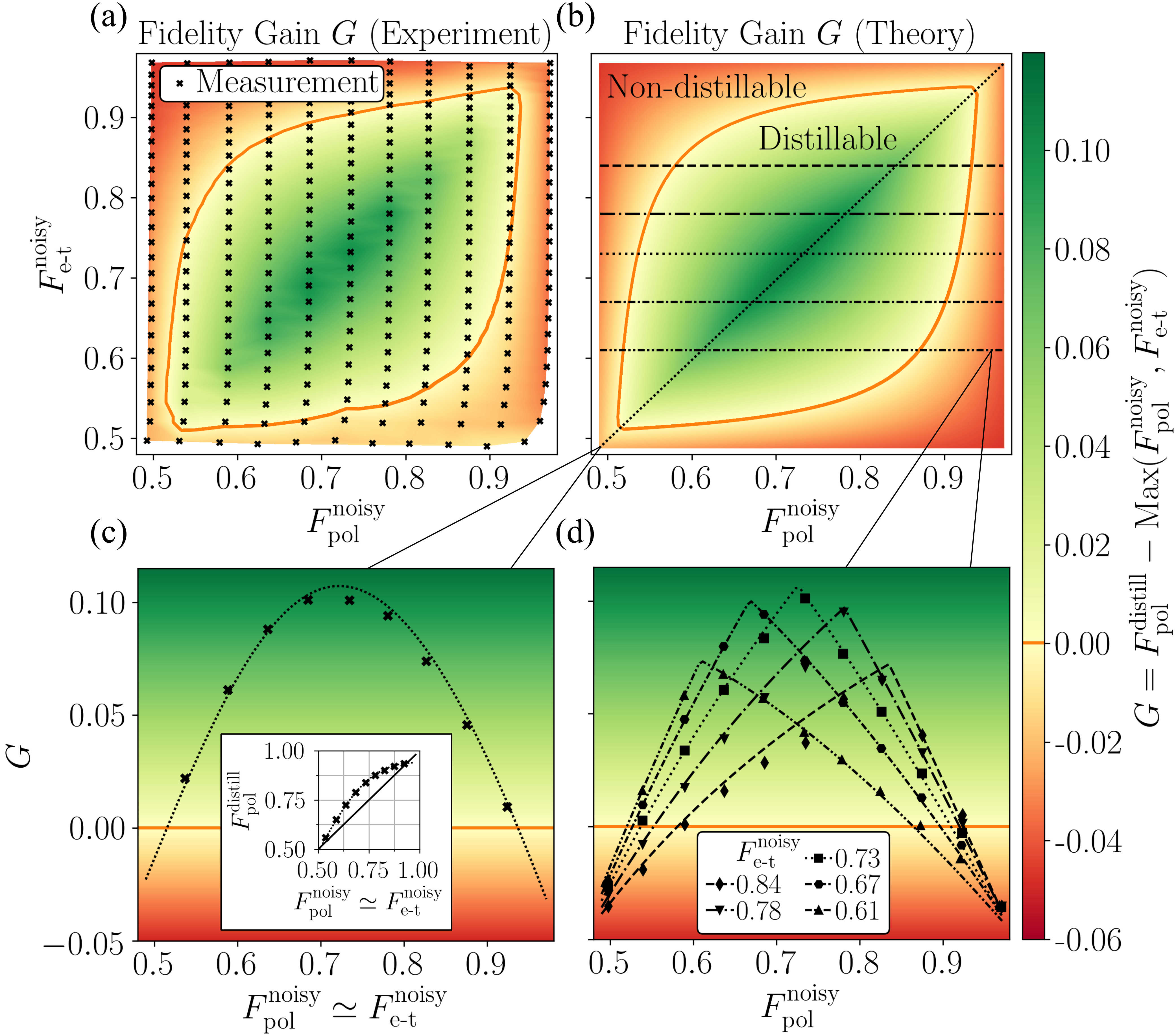} 
    \caption{
    Fidelity gain after single-copy entanglement distillation in polarisation and energy-time. 
    (a) The gain of the experimental data points are triangulated to form the heat map. Starting from an initial fidelity to the $\Phi^+$ state of $F^{\text{init}}_\text{pol} = \unit[97.1]{\%}$ and $F^{\text{init}}_\text{e-t} = \unit[96.8]{\%}$ (top right measurement point) we gradually increase the bit-flip (bit and bit-phase-flip) error in polarisation (energy-time) down to a fidelity of $\sim\unit[50]{\%}$. 
    (b) The heat map corresponds to the model, with the initial fidelities and the imperfect CNOT unitary as the only model parameters. 
    (c) A cut through the heat map at  $F^{\text{noisy}}_{\text{pol}}\simeq F^{\text{noisy}}_{\text{e-t}}$ reveals the characteristic behaviour of two-copy distillation \cite{bennett96}. 
    (d) Cuts through the heat map at constant $F^{\text{noisy}}_\text{e-t}$. 
    Both in  (c) and (d) lines correspond to the model, while markers correspond to measurement points with a standard deviation smaller than the marker size. The orange line separates the region of distillable ($G>0$, green) from the region of non-distillable states ($G<0$, red).
    }
    \label{fig:results}
\end{figure}

In order to fully characterise the performance of the protocol, we separately introduce different error types in the polarisation channel and the energy-time channel.
All trace-preserving errors on a qubit can be decomposed into the so-called error basis.
Apart from the identity $\sigma_0$, this operator basis consists of the Pauli operators $\sigma_\text{x}$ (bit-flip error), $\sigma_\text{z}$ (phase-flip error) and $\sigma_\text{y}$ (bit-phase-flip error), corresponding to rotations about the three Bloch sphere axes.
The error basis is therefore capable of transforming the $\Phi^+$ Bell state into any mixture of Bell states (``Bell-diagonal states'') by means of one-sided transformations.
We utilise this fact in the polarisation domain by inserting four waveplates in Bob's channel (``Noisy polarisation channel''), which introduce arbitrary superpositions in the error basis, and we generate mixed states by averaging over different waveplate settings. The noisy polarisation channel therefore results in a mixture of the $\Phi^+_\text{pol}$ state and an erroneous state $\rho^\text{err}_\text{pol}$.
We implement the noisy channel for the energy-time domain by gradually changing the coincidence window, leading to a mixture of the $\Phi^+_\text{e-t}$ state and an erroneous state $\rho_\text{e-t}^\text{err} =   (\ket{\Psi^+}\bra{\Psi^+}_\text{e-t} + \ket{\Psi^-}\bra{\Psi^-}_\text{e-t})/2$ (see Appendix Sec.~\ref{sec:infl-coinc-wind} and Sec.~\ref{sec:engery-time-noise}  for details on the noise control).
Owing to the independent manipulation of the polarisation and the energy-time domain, the resulting family of mixed states is still a product state $\rho_\text{pol}^\text{noisy} \otimes \rho_\text{e-t}^\text{noisy}$, with
\begin{align*}
    \rho_\text{pol}^\text{noisy} &= F^{\text{noisy}}_{\text{pol}} \ket{\Phi^+}\bra{\Phi^+}_\text{pol} + (1-F^{\text{noisy}}_{\text{pol}})\rho_\text{pol}^\text{err},\\
    \rho_\text{e-t}^\text{noisy} &= F^{\text{noisy}}_{\text{e-t}} \ket{\Phi^+}\bra{\Phi^+}_\text{e-t} + (1-F^{\text{noisy}}_{\text{e-t}})\rho_\text{e-t}^\text{err}.
\end{align*}
As we are now able to experimentally produce mixed entangled states, we employ them as input to our distillation procedure.
The success of the distillation protocol can be evaluated by comparing the noisy and distilled state with each other.
For this purpose, we utilise the quantum state fidelity $F$ to the $\Phi^+$ state before and after the distillation.
The fidelity is an easily measurable quantity and a good estimator of entanglement \cite{friis19}. At first, we measure the fidelity $F_\text{pol}^\text{noisy}=\bra{\Phi^+}\rho_\text{pol}^\text{noisy}\ket{\Phi^+}_\text{pol}$ of the noisy polarisation state $\rho_\text{pol}^\text{noisy}$ to the $\Phi^+_\text{pol}$ state and similarly the fidelity $F_\text{e-t}^\text{noisy}$ of the noisy energy-time state $\rho_\text{e-t}^\text{noisy}$. 
After the distillation step 
\begin{equation*}
        \rho_\text{pol}^\text{noisy} \otimes \rho_\text{e-t}^\text{noisy} \xrightarrow{\text{distillation}} \rho_\text{pol}^\text{distill},
\end{equation*}
the fidelity $F_\text{pol}^\text{distill}=\bra{\Phi^+}\rho_\text{pol}^\text{distill}\ket{\Phi^+}_\text{pol}$ of the distilled  state $\rho_\text{pol}^\text{distill}$ to the $\Phi^+_\text{pol}$ state determines whether the distillation was successful.
Since the single-copy scheme inherently consists of two independent error channels, we make use of the gain ${G = F^{\text{distill}}_{\text{pol}}-\text{Max}(F^{\text{noisy}}_{\text{pol}},F^{\text{noisy}}_{\text{e-t}})}$ as our figure of merit.
Experimentally, the fidelity $F$ to the $\Phi^+$ state is obtained by measuring the interference visibility $V_{ii} = \langle \sigma_i \otimes \sigma_i\rangle$ in three mutually unbiased bases $\{\sigma_\text{x}, \sigma_\text{y}, \sigma_\text{z}\}$, as  ${F =(1+V_\text{xx}-V_\text{yy}+V_\text{zz})/4}$.

We first introduce a bit-flip error (${\rho_\text{pol}^\text{err}=\ket{\Psi^+}\bra{\Psi^+}_\text{pol}}$) to the polarisation domain and compare the gain of the experimental data [Fig.~\hyperref[fig:results]{\ref*{fig:results}(a)}] with the theory prediction [Fig.~\hyperref[fig:results]{\ref*{fig:results}(b)}].
Our experimentally obtained fine-grained error map in Fig.~\ref{fig:results} allows us to analyse the performance of the distillation protocol in great detail.
Firstly, the region of positive gain ($G>0$) is approximately symmetrical about the diagonal line defined by  $F^{\text{noisy}}_{\text{pol}}\simeq F^{\text{noisy}}_{\text{e-t}}$. All distilled states within this region have a higher fidelity to the maximally entangled $\Phi^+_\text{pol}$ state than both noisy states before the distillation.
The non-distillable states outside of this area ($G<0$) are highly asymmetric in their error contribution in polarisation and energy-time, while the highest measured gains, up to $G=0.101$ (\unit[13.8]{\%} relative gain), can be observed for approximately symmetric error contributions ($F^{\text{noisy}}_{\text{pol}} = 0.734,  F^{\text{noisy}}_{\text{e-t}} = 0.732$).
A cross-section along the diagonal [Fig.~\hyperref[fig:results]{\ref*{fig:results}(c)}] reveals the gain curve known from a bit-flip channel in two-copy entanglement distillation \cite{bennett96} and indicates, in addition to cross-sections at constant energy-time fidelity [Fig.~\hyperref[fig:results]{\ref*{fig:results}(d)}], a good agreement between our experimental data and the theory. 
\begin{figure}[t]
    \centering
    \includegraphics[width=1\columnwidth]{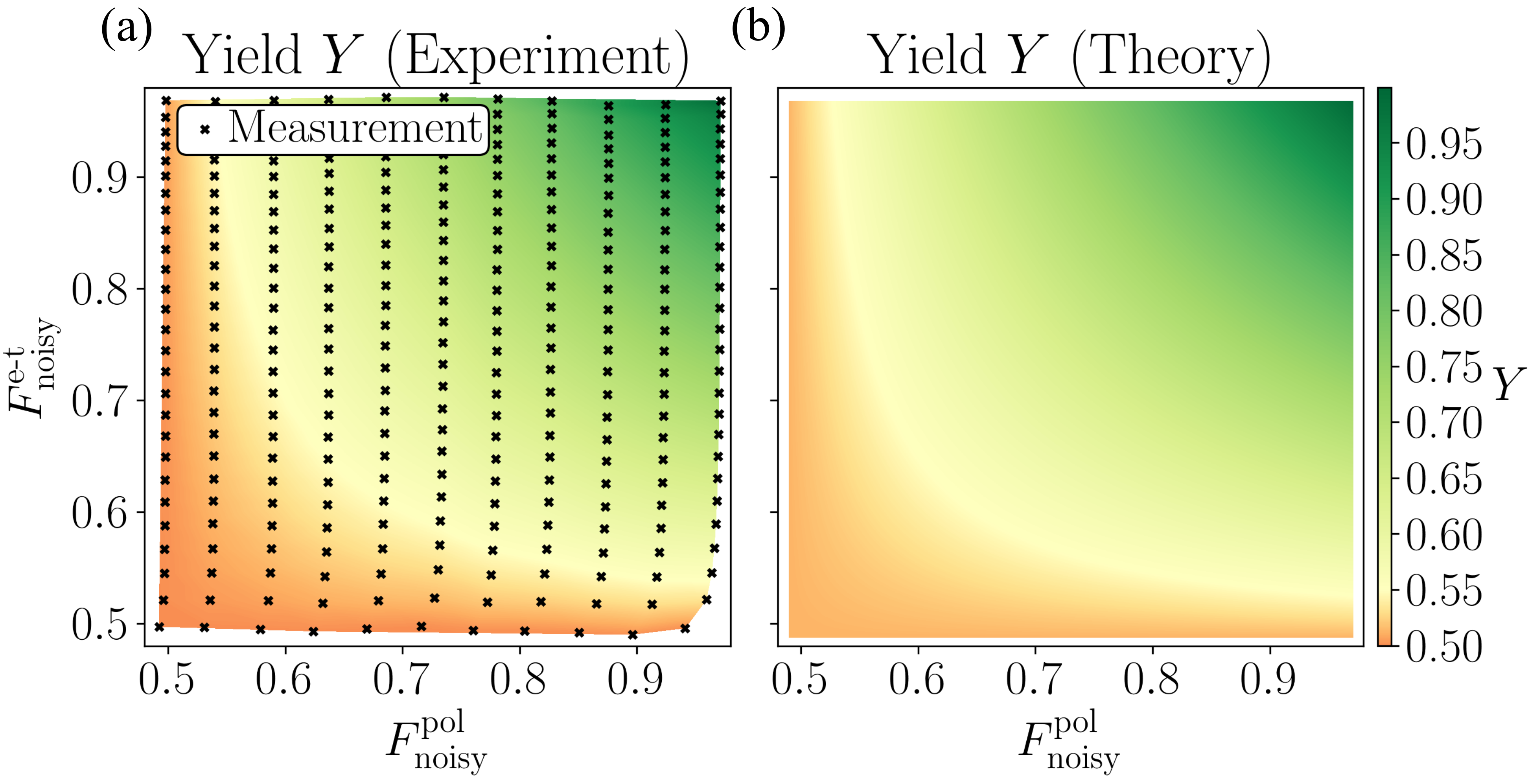}
    \caption{Yield of single-copy entanglement distillation in polarisation and energy-time. (a) The yield $Y$ corresponds to the measurement in Fig.~\ref{fig:results}. The heat maps are triangulations of (a) the measurement points and (b) the model data points.
    }
    \label{fig:yield}
\end{figure}
Another important figure of merit for distillation protocols is the yield.
For a single step of the protocol, we define the yield $Y$ as the number of distilled photon pairs divided by the total number of coincident photon pairs.
In our experiment, the yield peaks at $Y=\unit[98.6]{\%}$ for states close to the ideal source state in Eq.~\eqref{eq:pure_state}, while it does not fall below $Y=\unit[50]{\%}$ for any noisy state (Fig.~\ref{fig:yield}).
This is an improvement by a factor of 2 compared to the two-copy distillation protocol, which by default has to sacrifice the target photon pair to herald a successful distillation.

In order to demonstrate the universality of our scheme, we examine the bit-phase-flip  error (${\rho_\text{pol}^\text{err}=\ket{\Psi^-}\bra{\Psi^-}_\text{pol}}$) as an additional error type. 
We experimentally observe that this error shows the same gain and yield characteristics as the bit-flip error, and we again obtain a maximal gain of $G = 0.094$ (\unit[12.76]{\%} relative gain) at approximately symmetric error contributions ($F^{\text{noisy}}_{\text{pol}} = 0.733,  F^{\text{noisy}}_{\text{e-t}} = 0.733$). 
Thus, our results show that single-copy distillation can deal with various noise scenarios.

\section*{Discussion}
We experimentally demonstrated an entanglement distillation scheme based on interference between the polarisation and the energy-time domain of a single photon pair.
Our work constitutes the first experimental realisation of entanglement distillation based on CNOT gates, as initially devised in Refs.~\cite{simon02,bennett96}, between these DOF.
By using degrees of freedom which have been widely tested in long-distance experiments, the scheme we introduced can directly be implemented over existing fibre and free-space links.
Our thorough analysis of different error types and strengths allowed us to characterise the domain of distillable states and quantify the fidelity gain.
These results are of vital importance for future implementations in which, in general, different DOF are faced with different errors during transmission.

Apart from the experimental feasibility of our approach, the major advantage rests upon its high efficiency as compared to two-copy distillation~\cite{pan03}.
For typical experimental parameters and link losses of 20 dB in both channels, the single-copy distillation rate outperforms the tow-copy scheme by eight orders of magnitude (see Appendix Sec.~\ref{sec:rate} for a detailed rate comparison).
This illustrates that both the creation and the distribution of multiple photon pairs is costly, while a single hyperentangled photon pair can equivalently be utilised for entanglement distillation using significantly less resources.
Additionally, hyperentanglement is naturally produced in SPDC~\cite{Barreiro05} and does not necessarily increase the complexity of photon pair sources, which is essential for the establishment of entanglement distribution infrastructure such as satellite-based entanglement sources~\cite{yin17}. Furthermore, our implementation using the energy-time DOF is more robust in out-of-the-laboratory implementations as compared to realisations using the path DOF \cite{hu21}, which additionally requires modifications of the SPDC source.

As opposed to the two-copy scheme, our experiment neither depends on temporal synchronisation between Alice and Bob at the order of the photon's correlation time~\cite{valivarthi16}, nor on the storage in a quantum memory~\cite{seri17}.
For quantum cryptography in the polarisation domain, our scheme offers the added advantage of path multiplexing, potentially enhancing the secure key rates significantly~\cite{pseiner2020}.
While in its first conception~\cite{bennett96,deutsch96}, entanglement distillation was thought of as an asymptotic procedure converging on perfectly entangled states, single-copy distillation schemes are of course inherently limited to a finite number of distillation steps given by the number of accessible DOF (in our case two). 
Although the observed characteristic of the fidelity gain is evidence of its suitability for recurrence protocols~\cite{bennett96a},
practical considerations make clear that for all known applications in quantum information processing, perfect Bell pairs are not required and usually, crossing a certain noise threshold is sufficient, e.g. for yielding nonzero key rates in quantum key distribution. As our experiment shows, this is where single-copy distillation shows its true power.
In low-noise settings, the fidelity gains are moderate, while large gains are obtained in a regime where it really matters---that of high noise. This enables a significant increase in state quality in noise-dominated scenarios, recovering the potential for quantum communication applications in regimes that would otherwise be unattainable.

Photonic hyperentanglement in polarisation and energy-time has already been successfully demonstrated over a free-space link~\cite{steinlechner17}, where environmental noise is one of the limiting factors. Our experiment realistically simulates local noise on the polarisation degree of freedom, such as noise from stress-induced polarisation changes in fibres \cite{treiber2009}. Additionally, embedding the polarisation state in a higher-dimensional state space of energy-time or spatial modes can lead to the dilution of noise induced by background light or accidental coincidences (isotropic noise)~\cite{ecker19}.
Thus, combining the noise advantages of higher-dimensional systems with our implemented distillation procedure constitutes a promising path forward.

The presented scheme is, in principle, not limited to photonic implementations, and might be extended to other systems featuring hyperentanglement such as atoms \cite{mehwish17} or ions \cite{hu10}.
A natural extension of distillation protocols in two DOF will be the exploitation of more photonic DOF, enabling multiple distillation steps on a single photon pair.
The only prerequisite for including additional entangled DOF is the existence of corresponding CNOT gates~\cite{brandt:20, fiorentino04, hu21}.
Another viable path is the employment of generalised CNOT gates in high-dimensional entanglement distillation~\cite{miguel-ramiro18}. 

Distributing entanglement between remote parties in the face of noise is an essential task in quantum information processing.
Here, we tackled the problem of inefficient entanglement distillation by exploiting hyperentanglement of a single copy instead of exploiting two copies of a photon pair.
Our single-copy entanglement distillation approach enables the distribution of high-fidelity entangled states at practical rates and can thus become a vital building block of a future quantum internet. 
\\\\
\section*{Acknowledgements}
We thank Jan Lang for his support in the photon pair source alignment. We acknowledge funding from the Austrian Science Fund (FWF) through the START project Y879-N27 and from the European  Unions  Horizon  2020  programme grant agreement No. 857156 (OpenQKD).

\section*{Author Contributions}
S.E., P.S. and R.U. conceived the project; S.E., P.S. and L.B. designed and developed the experiment under the guidance of M.B. and R.U.; S.E., P.S. and L.B. evaluated the experimental data;  S.E. wrote the first draft of the manuscript; all authors discussed the results, contributed to writing and reviewed the manuscript; M.B. and R.U. supervised the project.

\clearpage
\onecolumngrid

\appendix
\renewcommand{\thesubsection}{\Alph{section}\arabic{subsection}}

\section*{Appendix}
	In this Appendix we describe our experimental methods, provide a detailed analysis of the modified Franson interferometer and illustrate the introduction of noise to the energy-time degree of freedom.
	The Appendix is organised as follows. In Sec.~\ref{sec:expmethods}, we describe our experimental setup and detail the employed methods. In Sec.~\ref{sec:work-princ-setup}, we discuss the working principle of the distillation setup and show how the two polarising beamsplitters act as a bilateral CNOT gate between the polarisation and the energy-time degree of freedom.
	The influence of the coincidence window is discussed in Sec.~\ref{sec:infl-coinc-wind}.
	In Sec.~\ref{sec:engery-time-noise}, we show how the error contribution in the energy-time degree of freedom can be controlled and how it affects the distillation.
	In Sec.~\ref{sec:rate}, we compare the distillation rate of the single-copy protocol with the two-copy protocol.


\section{Experimental Methods}
\label{sec:expmethods}
\subsection{Entangled photon pair source}
We produce photon pairs in spontaneous parametric down-conversion (SPDC) with type-II phase-matching by tightly focusing a continuous-wave pump beam into a 30-mm-long periodically-poled potassium titanyl phosphate (ppKTP) crystal.
The ppKTP crystal is temperature-tuned for degenerate phase-matching, yielding a centre wavelength of \unit[809.1]{nm} for the down-converted photons.
The pump laser (Toptica DL pro) is externally frequency-stabilised by locking it to a hyperfine transition of ${}^{39}\text{K}$, utilising  Doppler-free spectroscopy, resulting in a wavelength stability of $\sim\unit[0.6]{fm/min}$ at a wavelength of $\unit[404.53]{nm}$.
Polarisation entanglement is obtained by bidirectionally pumping the ppKTP crystal within a polarisation Sagnac interferometer. The entangled polarisation state can be controlled by a series of waveplates in the pump beam.
The photons are detected with avalanche photodiodes (Excelitas Technologies SPCM-800-11-FC) and each detection event is timestamped with a time-tagger (Swabian Instruments - Time Tagger Ultra).
A high degree of polarisation entanglement is accomplished by a good spatiotemporal overlap of the two SPDC processes, with typical interference visibilities of $V_\text{H/V}=\unit[99.9]{\%}$ and $V_\text{D/A}=\unit[97.8]{\%}$ in the rectilinear and diagonal polarisation basis, respectively.
After single-mode coupling we observe a photon-pair rate of \unit[68]{kcps per mW} of pump power at a symmetric heralding efficiency of $\unit[\sim 24]{\%}$. 

\subsection{Franson interferometer}
The two imbalanced Mach-Zehnder interferometers are stabilised by the frequency-locked pump laser.
Active phase stabilisation is achieved by a proportional–integral–derivative (PID) control loop of the difference signal from two photodiodes and a piezoactuator displacing both long arm mirrors. The stabilisation laser is injected into the unused port of the beam splitter and retrieved by a dichroic mirror after the interferometer, where the polarisation contrast is measured by fast photodiodes in a polarisation basis conjugate to the interferometer-defined polarisation basis. Phase-control is accomplished by tilting a multi-order waveplate in the pump beam, forcing the control loop to follow its setpoint. The time delay between the long and the short arm of the Mach-Zehnder interferometer is \unit[2.6]{ns}, which is larger than the FWHM  timing-jitter of the detection system ($\sim \unit[800]{ps}$)   and smaller than the coherence time of the pump laser ($\unit[\sim 600]{ns}$).

\subsection{Noisy channel}
In polarisation, we introduce errors by a sequence of four waveplates (``Noisy polarisation channel''). The relative rotation angles of the waveplates both determine the error type and the error contribution. 
We generate mixed entangled polarisation states by alternately setting the corresponding waveplate of the noisy polarisation channel to $+\theta$ and $-\theta$, with the neutral angle at $\theta=0$, leading to a reduction of coherences in the resulting state.
The controlled introduction of errors in the energy-time domain is more involved due to the lack of easy-to-implement unitaries.
We make use of the non-interfering states $\ket{t_\text{L},t_\text{S}}\bra{t_\text{L},t_\text{S}}$ and $\ket{t_\text{S},t_\text{L}}\bra{t_\text{S},t_\text{L}}$, which are usually discarded in coincidence postselection.
These admixtures correspond to a combined bit- and bit-phase-flip error $(\ket{\Psi^+}\bra{\Psi^+}_\text{e-t} + \ket{\Psi^-}\bra{\Psi^-}_\text{e-t})/2$. 
The error proportion can be tuned by gradually changing the coincidence window duration. 

\subsection{Fidelity measurements}
Different fidelity measurements require minor changes to the detection part of the setup. 
For $F_\text{pol}^\text{noisy}$ we insert the polariser (``Noisy state projection'') in front of the interferometer and block the long arm to prevent interference. 
The polarisers after the interferometer (``Distilled state projection'') are set to maximal transmission. The two linear polarisation bases \{H,V\} and \{D,A\} ($\sigma_\text{z} $ and $ \sigma_\text{x}$) can now be accessed by rotating the polariser. 
After inserting a quarter wave plate in front of the polariser, the circular polarisation basis \{R,L\} ($\sigma_\text{y}$) can be accessed. 
For the measurement of $F_\text{e-t}^\text{noisy}$, the long arm is unblocked. We measure in the computational basis $\{t_\text{L},t_\text{S}\}$ ($\sigma_\text{z}$) by setting both the noisy and distilled state projection polarisers at H-polarisation. 
Accessing the superposition bases $\{t_\text{L}\pm t_\text{S}\}$ and $\{t_\text{L}\pm i  t_\text{S}\}$ ($\sigma_\text{x} $ and $ \sigma_\text{y}$) is accomplished by setting both the noisy and distilled state projection polarisers at D-polarisation. 
The phase corresponding to each of the two bases is set via the phase control waveplate. 
The distilled fidelity $F_\text{pol}^\text{distill}$ is measured by removing the noisy state projection polarisers entirely. 
In total, each measurement point in main text Fig.~3 requires 54 different measurement settings, with an integration time of \unit[10]{s} per setting, amounting to 594 settings overall.
The coincidence rates at different coincidence windows are computed in post-processing. 
For large coincidence windows, the polarisation fidelity  $F_\text{pol}^\text{noisy}$ seems to get worse, especially for high fidelities (see main text Fig.~3a). 
This is due to a back-reflection during single-mode coupling in the photon pair source. 
The back-reflected photons make one additional round trip in the Sagnac interferometer, which agrees with the relative time delay we observe.

\section{Working principle of the setup and bilateral CNOT gate}
\label{sec:work-princ-setup}

\begin{figure}
	\centering
	\includegraphics[width=0.8\textwidth]{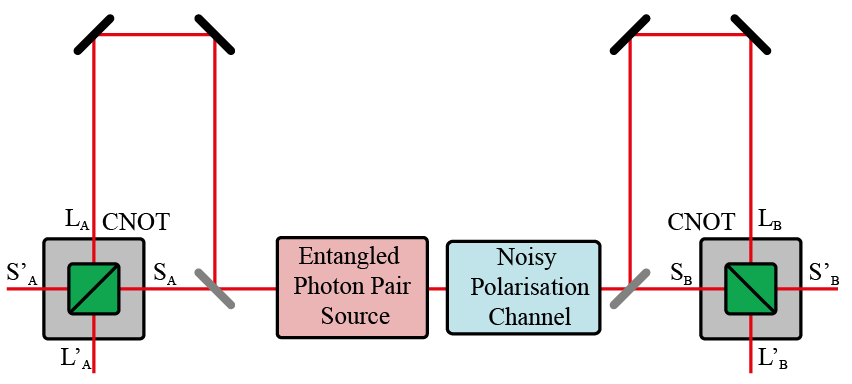}
	\caption{Simplified setup scheme.
		The basic structure of our distillation setup is a modified Franson interferometer~\cite{franson89}.
		While the input BS remain, the output BS are replaced by PBS as compared to the original scheme.
		The short paths in the unbalanced Mach-Zehnder interferometers are labelled with $\mathrm{S}_{\mathrm{A}}$ and $\mathrm{S}_{\mathrm{B}}$ at Alice's and Bob's side respectively.
		The long paths are labelled with $\mathrm{L}_{\mathrm{A}}$ and $\mathrm{L}_{\mathrm{B}}$.
		The modes transmitted by the PBS are marked with a dash.
		Locally, each PBS acts as CNOT gate with the polarisation DOF as control and the e-t (path) DOF as target.
		The polarisation mode $\mathrm{H}$ can be associated with the logical state $0$ and the polarisation mode $\mathrm{V}$ can be associated with the logical state $1$, i.e., the path mode is flipped if the photon incident on the PBS is vertically polarised and it remains unchanged if the photon is horizontally polarised.
		The joint action of both polarising beam splitters is a bilateral CNOT.
		(%
		BS: beam splitter,
		CNOT: controlled NOT,
		DOF: degree of freedom,
		PBS: polarising beam splitter%
		)
	}
	\label{fig:cw_simplified_setup}
\end{figure}

The energy-time~(e-t) degree of freedom~(DOF) distributed from the entangled-photon-pair source is analysed in a two-dimensional subspace by a Franson-type interferometer sketched in Figure~\ref{fig:cw_simplified_setup}.
Each balanced beamsplitter~(BS) splits one input mode into two path modes, which serve as input modes for the polarising BS~(PBS), closing the imbalanced Mach-Zehnder interferometer.
The two paths are not only characterised by their length $\mathrm{L}_{\mathrm{A}}$ and $\mathrm{L}_{\mathrm{B}}$, but also by the propagation time of the photons $\mathrm{t}_{\mathrm{L}_{\mathrm{A}}}$ and $\mathrm{t}_{\mathrm{L}_{\mathrm{B}}}$, such that e.g., $\mathrm{L}_{\mathrm{A}} = c\,\mathrm{t}_{\mathrm{L}_{\mathrm{A}}}$,  with the speed of light $c$.
Hence, there is a direct relation between the path and the e-t DOF.
In the following, we will use the path labels, while still referring to the e-t DOF.
When discussing the action of the PBS, we will keep track of the time separately.

The path length difference in each Mach-Zehnder interferometer, that is $\Delta\mathrm{L}_{\mathrm{A}} \equiv \mathrm{L}_{\mathrm{A}} - \mathrm{S}_{\mathrm{A}}$ at Alice's~(A) side and $\Delta\mathrm{L}_{\mathrm{B}} \equiv \mathrm{L}_{\mathrm{B}} - \mathrm{S}_{\mathrm{B}}$ at Bob's side, introduces a time delay. 
These time delays should be well between the single-photon coherence time~$\tau_\mathrm{sp}$ and the coherence time of the pump laser~$\tau_\mathrm{pump}$
\begin{align}
	\label{eq:MZ-time-delay-single-photon}
	\tau_\mathrm{sp} \ll \frac{\Delta \mathrm{L}_{i}}{\mathrm{c}} \ll \tau_\mathrm{pump}\,,
\end{align}
with $i\in\left\lbrace\mathrm{A,B}\right\rbrace$.
The lower limit on the imbalance ensures that no single photon interference is observed locally at A or B, while the upper limit on the imbalance ensures the coherence of the possible two-photon modes $\mathrm{S}_{\mathrm{A}}\mathrm{S}_{\mathrm{B}}$, $\mathrm{S}_{\mathrm{A}}\mathrm{L}_{\mathrm{B}}$, $\mathrm{L}_{\mathrm{A}}\mathrm{S}_{\mathrm{B}}$ and $\mathrm{L}_{\mathrm{A}}\mathrm{L}_{\mathrm{B}}$.
If further the interferometer imbalances at A and B are the same $\left(\Delta\mathrm{L}_{\mathrm{A}} = \Delta\mathrm{L}_{\mathrm{B}}\right)$, the modes $\mathrm{S}_{\mathrm{A}}\mathrm{S}_{\mathrm{B}}$ and $\mathrm{L}_{\mathrm{A}}\mathrm{L}_{\mathrm{B}}$ are indistinguishable by their relative time delay $t_\text{A} - t_\text{B} \equiv 0$.
The remaining modes $\mathrm{S}_{\mathrm{A}}\mathrm{L}_{\mathrm{B}}$ and $\mathrm{L}_{\mathrm{A}}\mathrm{S}_{\mathrm{B}}$, on the contrary, are clearly distinguishable by their time delays $t_\text{A} - t_\text{B} \equiv -\Delta T$ and $\Delta T$ respectively.
In a delay histogram as shown in Fig.~\hyperref[fig:cw_histo]{\ref*{fig:cw_histo}a}, they form the side-peaks centred around the central-peak at time delay 0.

\section{Influence of the coincidence window}
\label{sec:infl-coinc-wind}

By choosing the time window for coincidence counting small enough, the side peaks can be excluded in post-processing, such that the Bell state $\ket{\Phi^+}_\mathrm{e-t} = \left(\ket{\mathrm{S}_{\mathrm{A}}\mathrm{S}_{\mathrm{B}}} + \ket{\mathrm{L}_{\mathrm{A}}\mathrm{L}_{\mathrm{B}}}\right)/\sqrt{2}$ can be prepared if the phases in the long paths are adjusted.
The $\Phi^+$ state is of particular interest, as the contribution of $\Phi^+$ states in both DOF remains unchanged under the joint action of the PBS.
As each PBS acts as a CNOT gate between the polarisation DOF and the e-t DOF, we will call the joint action of both PBS a bilateral CNOT~(bCNOT).

\begin{figure}
	\centering
	\includegraphics[width=0.8\textwidth]{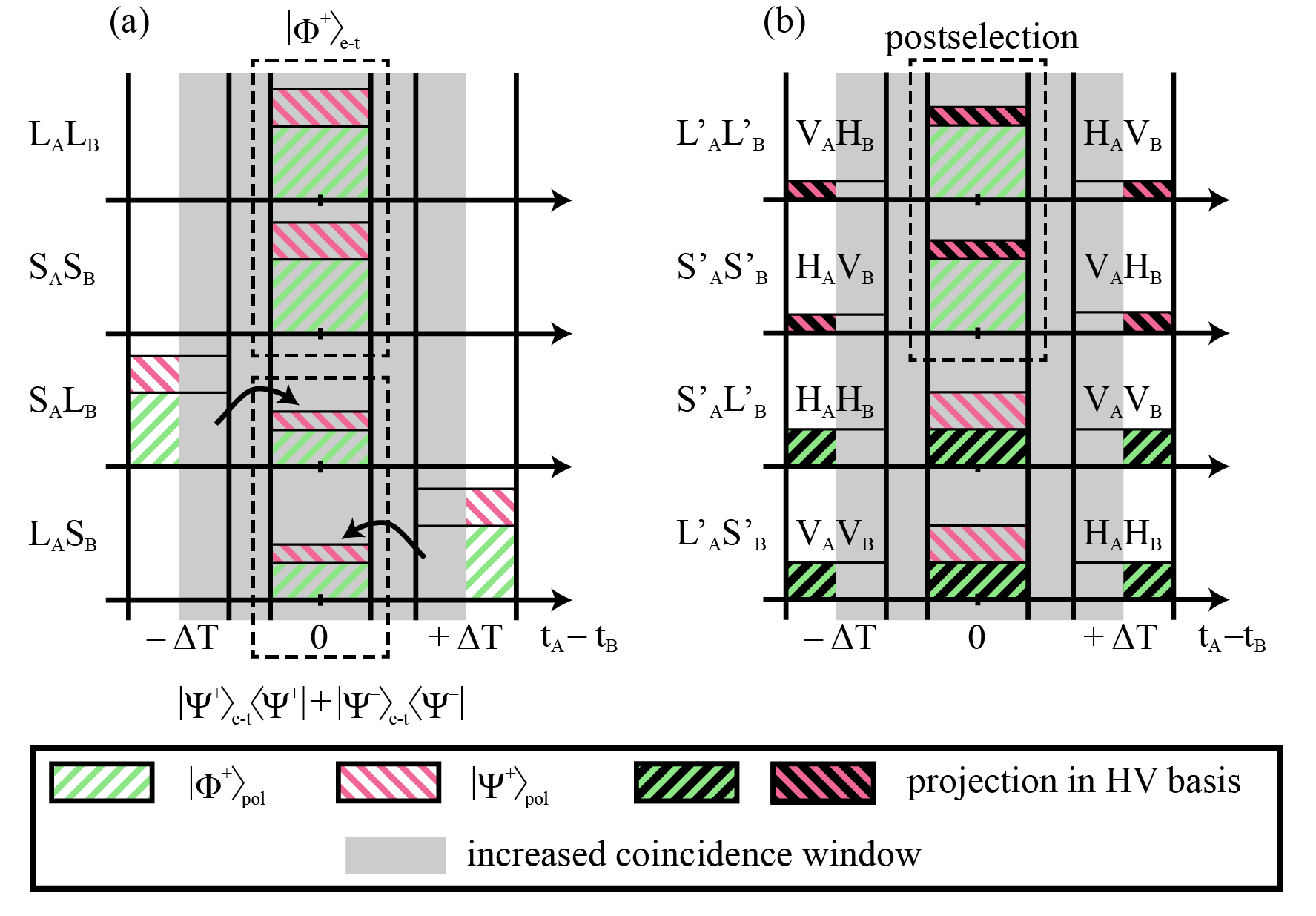}
	\caption{
		Delay histogram visualising the effect of an increased coincidence window on the two-photon state in both the polarisation and the energy time DOF before~(a) and after~(b) the bCNOT.
		The state in the e-t DOF is depicted by 12 slots spanned by three orthogonal states with relative time delays $-\Delta T$, $0$ and $\Delta T$ and four orthogonal path states.
		The polarisation state is depicted by hatched regions, where the proportion of the regions represent the proportion of the contributions to the mixed state.
		We consider the target state $\Phi^+_\mathrm{pol}$ with an admixed bit flip contribution $\Psi^+_\mathrm{pol}$.
		The initial amount of bit flip error in polarisation as well as the width of the coincidence window are chosen arbitrarily.
		After the distillation (b), which includes the action of the bCNOT as well as the postselection on the $\Phi^\pm_\mathrm{e-t}$ states in the e-t DOF, the unwanted contributions are clearly reduced compared to the input polarisation state before the bCNOT~(a).
		A detailed description of the state evolution is given in the text.
		(%
		bCNOT: bilateral controlled NOT,
		DOF: degree of freedom%
		)\\
	}
	\label{fig:cw_histo}
\end{figure}

In Figure~\hyperref[fig:cw_histo]{\ref*{fig:cw_histo}a}, the effect of an increased coincidence window on the state before the distillation step is illustrated.
If the coincidence window partially includes the side-peaks, both the central-peak, as well as parts of the side-peak are identified as coincidence by the logic.
However, since contributions from the side-peaks are in principle physically distinguishable by their relative time delay, they form a non-interfering background.
In case of an equal contribution of the side-peaks, the non-interfering background is proportional to a balanced mixture of the $\Psi^\pm_\mathrm{e-t}$ states

\begin{align}
	\label{eq:non-interfering-background}
	\kb{\mathrm{S}_{\mathrm{A}}\mathrm{L}_{\mathrm{B}}} + \kb{\mathrm{L}_{\mathrm{A}}\mathrm{S}_{\mathrm{B}}} = \kbsub{\Psi^+}{\mathrm{e-t}} + \kbsub{\Psi^-}{\mathrm{e-t}}\,.
\end{align}

\begin{figure}
	\centering
	\includegraphics[width=0.8\textwidth]{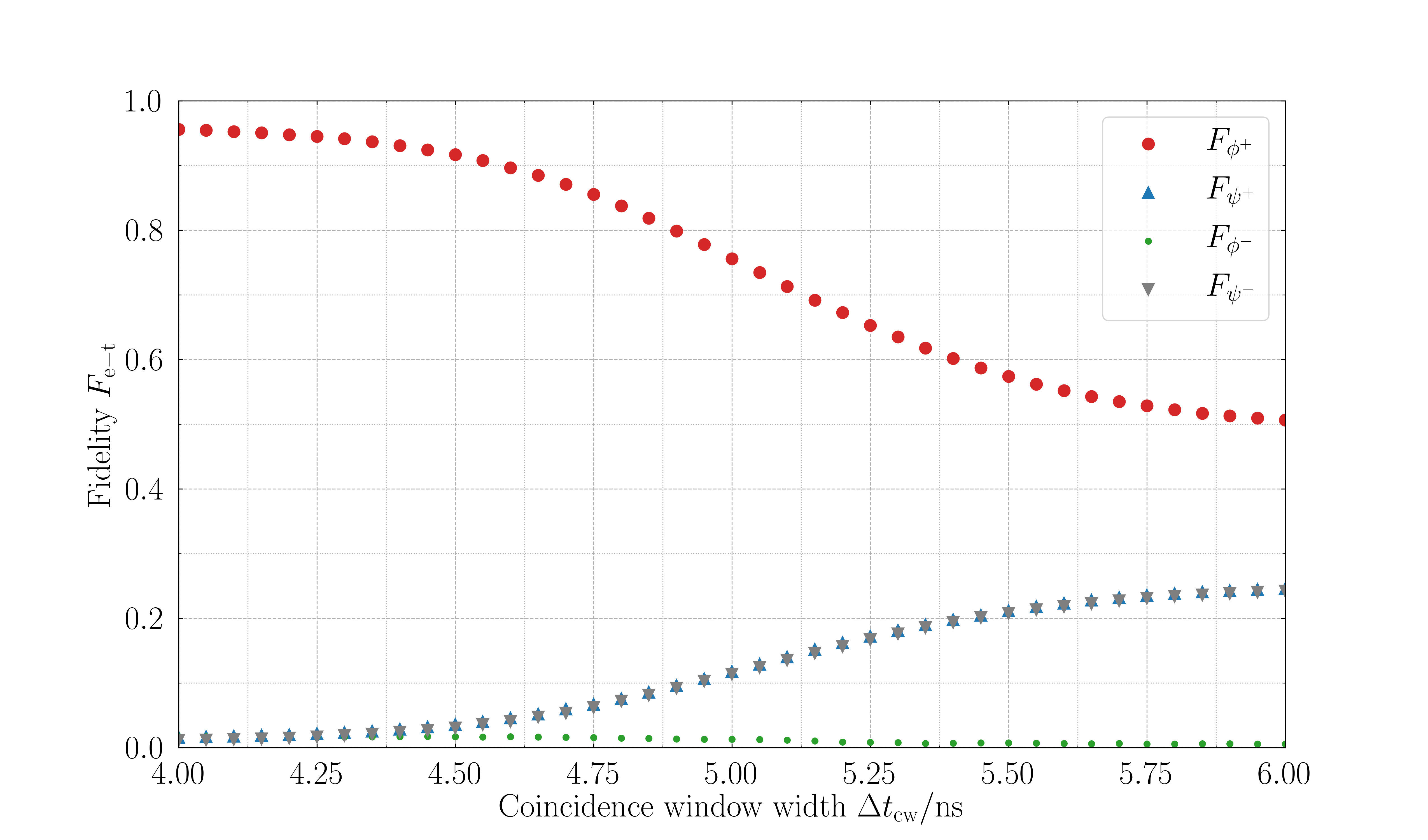}
	\caption{Experimental data of the Bell state fidelities in the e-t DOF before the bCNOT as a function of the coincidence window width.
		With an increasing coincidence window, the fidelities to the $\Psi^\pm$ states rise equally up to $0.25$ once the side-peaks are fully included in the coincidence window.
		At the same time, the fidelity to the target state of the distillation process $\Phi^+$ declines to $0.5$.
		The fidelity to the $\Psi^-$ state is non-zero due to residual phase errors.
		The error bars are too small to be shown.
		(%
		bCNOT: bilateral controlled NOT,
		DOF: degree of freedom,
		e-t: energy-time%
		)
	}
	\label{fig:cow_fid}
\end{figure}

In Figure~\ref{fig:cow_fid} we gradually increased the coincidence window of experimental data, and indeed found an increasing admixture of the states in Eq.~\eqref{eq:non-interfering-background}. 
While the illustration by the histogram~(see Fig.~\eqref{fig:cw_histo}) might give the impression that there is a linear relation between the coincidence window width and the Bell state fidelity in the e-t DOF, the experimental data show that this is not the case.
Instead, the experimental data reflects the Gaussian shape of the coincidence peaks governed by the timing jitter of the photon detectors.

\section{Energy-time noise in the distillation protocol}
\label{sec:engery-time-noise}

So far, the polarisation DOF did not play a role.
This changes once we consider the effect of the bCNOT, as it is implemented by two polarising beam splitters.
Each PBS acts as CNOT between with the polarisation DOF as control qubit and the path DOF as target qubit, e.g.,
\begin{align}
	\label{eq:PBS-CNOT}
	\ket{\mathrm{H}}_\mathrm{pol}\otimes\ket{\mathrm{S}_{\mathrm{A}}}_\mathrm{e-t} &\xrightarrow[]{\mathrm{CNOT}} \ket{\mathrm{H}}_\mathrm{pol}\otimes\ket{\mathrm{S}_{\mathrm{A}}'}_\mathrm{e-t}\,,\\
	\ket{\mathrm{H}}_\mathrm{pol}\otimes\ket{\mathrm{L}_{\mathrm{A}}}_\mathrm{e-t} &\xrightarrow[]{\mathrm{CNOT}} \ket{\mathrm{H}}_\mathrm{pol}\otimes\ket{\mathrm{L}_{\mathrm{A}}'}_\mathrm{e-t}\,,\\
	\ket{\mathrm{V}}_\mathrm{pol}\otimes\ket{\mathrm{S}_{\mathrm{A}}}_\mathrm{e-t} &\xrightarrow[]{\mathrm{CNOT}} \ket{\mathrm{V}}_\mathrm{pol}\otimes\ket{\mathrm{L}_{\mathrm{A}}'}_\mathrm{e-t}\,,\\
	\ket{\mathrm{V}}_\mathrm{pol}\otimes\ket{\mathrm{L}_{\mathrm{A}}}_\mathrm{e-t} &\xrightarrow[]{\mathrm{CNOT}} \ket{\mathrm{V}}_\mathrm{pol}\otimes\ket{\mathrm{S}_{\mathrm{A}}'}_\mathrm{e-t}\,.
\end{align}
While the PBS swap the path mode depending on the polarisation state, they do not affect the time delays.
In the delay histogram~(see Fig.~\ref{fig:cw_histo}), this means that contributions are only shifted vertically (path domain), not horizontally (time domain).
Unlike the path labels before the bCNOT, the dashed path labels $\mathrm{S}_{\mathrm{A}}'$, $\mathrm{L}_{\mathrm{A}}'$, $\mathrm{S}_{\mathrm{B}}'$ and $\mathrm{L}_{\mathrm{B}}'$ after the bCNOT cannot be directly identified with a time delay, e.g. a photon detected in mode $\mathrm{S}_{\mathrm{A}}'$ can, depending on the polarisation, either have been transmitted from path $\mathrm{S}_{\mathrm{A}}$ or reflected from path $\mathrm{L}_{\mathrm{A}}$.
For the path modes of the two photons, this means that while a photon pair in the modes $\mathrm{S}_{\mathrm{A}}$ and $\mathrm{S}_{\mathrm{B}}$ can be unambiguously assigned to the central-peak, a photon pair in the modes $\mathrm{S}_{\mathrm{A}}'$ and $\mathrm{S}_{\mathrm{B}}'$ can likewise originate from one of the side-peaks.

Let us now have a closer look on how the state illustrated in the histogram in Figure~\hyperref[fig:cw_histo]{\ref*{fig:cw_histo}a} evolves by the action of the bCNOT to the state illustrated by the histogram in Figure~\hyperref[fig:cw_histo]{\ref*{fig:cw_histo}b}.
The polarisation state $\Phi^+_\mathrm{pol}$ with a bit-flip admixture $\Psi^+_\mathrm{pol}$ can be described by the Bell-diagonal density matrix
\begin{align}
	\label{eq:dm-pol}
	\rho_\mathrm{pol} = F_{\mathrm{pol}} \kbsub{\Phi^+}{\mathrm{pol}} + (1-F_{\mathrm{pol}})\kbsub{\Psi^+}{\mathrm{pol}}\,,
\end{align}
where $F_{\mathrm{pol}}$ is the polarisation fidelity.
As demonstrated in Equation~(\ref{eq:non-interfering-background}), an increased coincidence window introduces equal admixtures of a bit-flip error $\Psi^+_\mathrm{e-t}$ and a bit-phase-flip error $\Psi^-_\mathrm{e-t}$ to the e-t state, which can be described by the Bell-diagonal density matrix
\begin{align}
	\label{eq:dm-et}
	\rho_\mathrm{e-t} = F_{\mathrm{e-t}} \kbsub{\Phi^+}{\mathrm{e-t}} + \frac{1-F_{\mathrm{e-t}}}{2}\left(\kbsub{\Psi^+}{\mathrm{e-t}} + \kbsub{\Psi^-}{\mathrm{e-t}}\right)\,,
\end{align}
with the e-t fidelity $F_\mathrm{e-t}$.
The tensor product of those two matrices $\rho = \rho_\mathrm{pol} \otimes \rho_\mathrm{e-t}$ describes the state of both DOF and has six non-zero entries in the Bell-diagonal form.
For simplicity, the action of the bCNOT will be discussed on each of these six contributions separately.

Beforehand, let us clarify the action of the bCNOT with the help of an example.
A photon pair in the $\ketab{\mathrm{H}}{\mathrm{V}}$ state incident on both PBS from the short path is transformed as
\begin{align}
	\label{eq:bCNOT-example}
	\ketab{\mathrm{H}}{\mathrm{V}} \otimes \ket{\mathrm{S}_{\mathrm{A}}\mathrm{S}_{\mathrm{B}}} \xrightarrow[]{\mathrm{bCNOT}} \ketab{\mathrm{H}}{\mathrm{V}} \otimes \ket{\mathrm{S}_{\mathrm{A}}'\mathrm{L}_{\mathrm{B}}'}\,.
\end{align}
The polarisation remains unchanged and the path mode is flipped only if a photon is vertically polarised.

The state with both DOF in the $\Phi^+$ state remains unchanged under the action of the bCNOT
\begin{subequations}
	\begin{alignat}{2}
		\label{eq:bCNOT-phip-interfering}
		\ket{\Phi^+}_\mathrm{pol} \otimes \ket{\Phi^+}_\mathrm{e-t}=
		&\frac{1}{2} \left[\ketab{\mathrm{H}}{\mathrm{H}} \otimes \left(\ket{\mathrm{S}_{\mathrm{A}}\mathrm{S}_{\mathrm{B}}} + \ket{\mathrm{L}_{\mathrm{A}}\mathrm{L}_{\mathrm{B}}}\right) +
		\ketab{\mathrm{V}}{\mathrm{V}} \otimes \left(\ket{\mathrm{S}_{\mathrm{A}}\mathrm{S}_{\mathrm{B}}} + \ket{\mathrm{L}_{\mathrm{A}}\mathrm{L}_{\mathrm{B}}}\right)\right]\\
		\xrightarrow[]{\mathrm{bCNOT}}
		&\frac{1}{2} \left[\ketab{\mathrm{H}}{\mathrm{H}} \otimes \left(\ket{\mathrm{S}_{\mathrm{A}}'\mathrm{S}_{\mathrm{B}}'} + \ket{\mathrm{L}_{\mathrm{A}}'\mathrm{L}_{\mathrm{B}}'}\right) +
		\ketab{\mathrm{V}}{\mathrm{V}} \otimes \left(\ket{\mathrm{L}_{\mathrm{A}}'\mathrm{L}_{\mathrm{B}}'} + \ket{\mathrm{S}_{\mathrm{A}}'\mathrm{S}_{\mathrm{B}}'}\right)\right]\\
		= &\ket{\Phi^+}_\mathrm{pol} \otimes \ket{\Phi^+}_\mathrm{e-t}\,.
	\end{alignat}
\end{subequations}
This property makes the $\Phi^+$ state the target state of the distillation scheme.
The erroneous contributions in polarisation paired with the $\Phi^+_\text{e-t}$ state in the e-t DOF are transformed as
\begin{subequations}
	\begin{alignat}{2}
		\label{eq:bCNOT-psip-interfering}
		\ket{\Psi^+}_\mathrm{pol} \otimes \ket{\Phi^+}_\mathrm{e-t}=
		&\frac{1}{2}\left[ \ketab{\mathrm{H}}{\mathrm{V}} \otimes \left(\ket{\mathrm{S}_{\mathrm{A}}\mathrm{S}_{\mathrm{B}}} + \ket{\mathrm{L}_{\mathrm{A}}\mathrm{L}_{\mathrm{B}}}\right) +
		\ketab{\mathrm{V}}{\mathrm{H}} \otimes \left(\ket{\mathrm{S}_{\mathrm{A}}\mathrm{S}_{\mathrm{B}}} + \ket{\mathrm{L}_{\mathrm{A}}\mathrm{L}_{\mathrm{B}}}\right)\right]\\
		\xrightarrow[]{\mathrm{bCNOT}}
		&\frac{1}{2} \left[\ketab{\mathrm{H}}{\mathrm{V}} \otimes \left(\ket{\mathrm{S}_{\mathrm{A}}'\mathrm{L}_{\mathrm{B}}'} + \ket{\mathrm{L}_{\mathrm{A}}'\mathrm{S}_{\mathrm{B}}'}\right) +
		\ketab{\mathrm{V}}{\mathrm{H}} \otimes \left(\ket{\mathrm{L}_{\mathrm{A}}'\mathrm{S}_{\mathrm{B}}'} + \ket{\mathrm{S}_{\mathrm{A}}'\mathrm{L}_{\mathrm{B}}'}\right)\right]\\
		=& \ket{\Psi^+}_\mathrm{pol} \otimes \ket{\Psi^+}_\mathrm{e-t}\,.
	\end{alignat}
\end{subequations}
After the bCNOT, these contributions are discarded by the postselection on the $\Phi^\pm_\text{e-t}$  states in energy-time.
The postselection projector reads
\begin{align}
	\label{eq:projector-phip-phim}
	\mathrm{P}_{\Phi^\pm} = \mathbbm{1}_\mathrm{pol} \otimes \left(\kbsub{\Phi^+}{\mathrm{e-t}} + \kbsub{\Phi^-}{\mathrm{e-t}}\right) = \mathbbm{1}_\mathrm{pol} \otimes \left(\kb{\mathrm{S}_{\mathrm{A}}'\mathrm{S}_{\mathrm{B}}'} + \kb{\mathrm{L}_{\mathrm{A}}'\mathrm{L}_{\mathrm{B}}'}\right)\,.
\end{align}

Similarly, the side-peaks are transformed as
\begin{align}
	\label{eq:bCNOT-phip-SL}
	&\ket{\Phi^+}_\mathrm{pol} \otimes \ket{\mathrm{S}_{\mathrm{A}}\mathrm{L}_{\mathrm{B}}} \xrightarrow[]{\mathrm{bCNOT}} \ketab{\mathrm{H}}{\mathrm{H}} \otimes \ket{\mathrm{S}_{\mathrm{A}}'\mathrm{L}_{\mathrm{B}}'} + \ketab{\mathrm{V}}{\mathrm{V}} \otimes \ket{\mathrm{L}_{\mathrm{A}}'\mathrm{S}_{\mathrm{B}}'}\,,\\
	\label{eq:bCNOT-phip-LS}
	&\ket{\Phi^+}_\mathrm{pol} \otimes \ket{\mathrm{L}_{\mathrm{A}}\mathrm{S}_{\mathrm{B}}} \xrightarrow[]{\mathrm{bCNOT}} \ketab{\mathrm{H}}{\mathrm{H}} \otimes \ket{\mathrm{L}_{\mathrm{A}}'\mathrm{S}_{\mathrm{B}}'} + \ketab{\mathrm{V}}{\mathrm{V}} \otimes \ket{\mathrm{S}_{\mathrm{A}}'\mathrm{L}_{\mathrm{B}}'}\,.
\end{align}
As the coincidence window is widened, the non-interfering background can be described as a mixture of Bell states as noted in Equation~(\ref{eq:dm-et}).
While the transformation of those contributions is completely described by Equations~\eqref{eq:bCNOT-phip-SL} and \eqref{eq:bCNOT-phip-LS}, it can be rephrased in the Bell basis as
\begin{align}
	\label{eq:bCNOT-phip-psip}
	&\ket{\Phi^+}_\mathrm{pol} \otimes \ket{\Psi^+}_\mathrm{e-t} \xrightarrow[]{\mathrm{bCNOT}} \ket{\Phi^+}_\mathrm{pol} \otimes \ket{\Psi^-}_\mathrm{e-t}\,,\\
	\label{eq:bCNOT-phip-psim}
	&\ket{\Phi^+}_\mathrm{pol} \otimes \ket{\Psi^-}_\mathrm{e-t} \xrightarrow[]{\mathrm{bCNOT}} \ket{\Phi^-}_\mathrm{pol} \otimes \ket{\Psi^-}_\mathrm{e-t}\,.
\end{align}
In this case, contributions from the target state in polarisation are discarded in postselection.
In contrast, erroneous polarisation contributions that are within the increased coincidence window, are kept in postselection.
The side-peaks transform like
\begin{align}
	\label{eq:bCNOT-psip-noninterfering}
	&\ket{\Psi^+}_\mathrm{pol} \otimes \ket{\mathrm{S}_{\mathrm{A}}\mathrm{L}_{\mathrm{B}}} \xrightarrow[]{\mathrm{bCNOT}} \ketab{\mathrm{H}}{\mathrm{V}} \otimes \ket{\mathrm{S}_{\mathrm{A}}'\mathrm{S}_{\mathrm{B}}'} + \ketab{\mathrm{V}}{\mathrm{H}} \otimes \ket{\mathrm{L}_{\mathrm{A}}'\mathrm{L}_{\mathrm{B}}'}\,,\\
	&\ket{\Psi^+}_\mathrm{pol} \otimes \ket{\mathrm{L}_{\mathrm{A}}\mathrm{S}_{\mathrm{B}}} \xrightarrow[]{\mathrm{bCNOT}} \ketab{\mathrm{H}}{\mathrm{V}} \otimes \ket{\mathrm{L}_{\mathrm{A}}'\mathrm{L}_{\mathrm{B}}'} + \ketab{\mathrm{V}}{\mathrm{H}} \otimes \ket{\mathrm{S}_{\mathrm{A}}'\mathrm{S}_{\mathrm{B}}'}\,,
\end{align}
which can be expressed for the contributions included in the coincidence window as
\begin{align}
	\label{eq:bCNOT-psip-psip}
	&\ket{\Psi^+}_\mathrm{pol} \otimes \ket{\Psi^+}_\mathrm{e-t} \xrightarrow[]{\mathrm{bCNOT}} \ket{\Psi^+}_\mathrm{pol} \otimes \ket{\Phi^+}_\mathrm{e-t}\,,\\
	\label{eq:bCNOT-psip-psim}
	&\ket{\Psi^+}_\mathrm{pol} \otimes \ket{\Psi^-}_\mathrm{e-t} \xrightarrow[]{\mathrm{bCNOT}} \ket{\Psi^-}_\mathrm{pol} \otimes \ket{\Phi^-}_\mathrm{e-t}\,.
\end{align}

We showed that widening the coincidence window does indeed have the desired effect of creating a mixed state in the energy-time degree of freedom.
Further, we discussed a distillation step including joint action of the polarising beams splitters as well as the postselection in detail and illustrated how the proportion of unwanted contribution to the target state in polarisation is thereby reduced.

\section{Comparison of distillation rates}
\label{sec:rate}

In the main text, we compare the single-copy distillation rate with the two-copy distillation rate~\cite{pan03}.
In Table~\ref{tab:dist_rate} we list the most relevant contributions to the distillation rate for the two protocols, starting from the photon-pair creation with pulsed SPDC sources and the subsequent transmission over a dual link, to the distillation operation itself.
Further rate-diminishing factors in the two-copy distillation protocol, such as inefficient quantum memories \cite{seri17} and inherently probabilistic CNOT gates \cite{gasparoni04}, are not included in this table.

\renewcommand{\arraystretch}{1.4}
\begin{table}[H]
	\caption{\label{tab:dist_rate} Comparison of distillation rates between single-copy and two-copy entanglement distillation. The pulsed SPDC source used in the two-copy distillation experiment \cite{pan03} produces a mean number of photon pairs per pulse of $p=\unit[0.00022]{}$ at a repetition rate of $R_\text{rep} = \unit[76]{MHz}$. We assume a link transmittance of $t=\unit[-20]{dB}$ and a yield of $Y=0.8$.}
	\begin{ruledtabular}
		\begin{center}
			\begin{tabular}{>{\centering\arraybackslash}m{3cm}>{\centering\arraybackslash}m{3cm}>{\centering\arraybackslash}b{4cm}>{\centering\arraybackslash}p{2cm}>{\centering\arraybackslash}p{5cm}}
				Distillation scheme & Creation probability & Transmission probability & Protocol yield & Distillation rate [1/$s$]  \\ \hline  
				Single copy & $p$ & $t^2$ & $Y$ & $R_\text{rep}pt^2Y$ = $1.36$ \\
				Two copy & $p^2$ & $t^4$ & $Y/2$ &       $ \;\;\;\;\;R_\text{rep}p^2t^4Y/2$ = $1.52\cdot10^{-8}$
			\end{tabular}
		\end{center}
	\end{ruledtabular}
\end{table}


\begin{thebibliography}{57}%
	\makeatletter
	\providecommand \@ifxundefined [1]{%
		\@ifx{#1\undefined}
	}%
	\providecommand \@ifnum [1]{%
		\ifnum #1\expandafter \@firstoftwo
		\else \expandafter \@secondoftwo
		\fi
	}%
	\providecommand \@ifx [1]{%
		\ifx #1\expandafter \@firstoftwo
		\else \expandafter \@secondoftwo
		\fi
	}%
	\providecommand \natexlab [1]{#1}%
	\providecommand \enquote  [1]{``#1''}%
	\providecommand \bibnamefont  [1]{#1}%
	\providecommand \bibfnamefont [1]{#1}%
	\providecommand \citenamefont [1]{#1}%
	\providecommand \href@noop [0]{\@secondoftwo}%
	\providecommand \href [0]{\begingroup \@sanitize@url \@href}%
	\providecommand \@href[1]{\@@startlink{#1}\@@href}%
	\providecommand \@@href[1]{\endgroup#1\@@endlink}%
	\providecommand \@sanitize@url [0]{\catcode `\\12\catcode `\$12\catcode
		`\&12\catcode `\#12\catcode `\^12\catcode `\_12\catcode `\%12\relax}%
	\providecommand \@@startlink[1]{}%
	\providecommand \@@endlink[0]{}%
	\providecommand \url  [0]{\begingroup\@sanitize@url \@url }%
	\providecommand \@url [1]{\endgroup\@href {#1}{\urlprefix }}%
	\providecommand \urlprefix  [0]{URL }%
	\providecommand \Eprint [0]{\href }%
	\providecommand \doibase [0]{http://dx.doi.org/}%
	\providecommand \selectlanguage [0]{\@gobble}%
	\providecommand \bibinfo  [0]{\@secondoftwo}%
	\providecommand \bibfield  [0]{\@secondoftwo}%
	\providecommand \translation [1]{[#1]}%
	\providecommand \BibitemOpen [0]{}%
	\providecommand \bibitemStop [0]{}%
	\providecommand \bibitemNoStop [0]{.\EOS\space}%
	\providecommand \EOS [0]{\spacefactor3000\relax}%
	\providecommand \BibitemShut  [1]{\csname bibitem#1\endcsname}%
	\let\auto@bib@innerbib\@empty
	\bibitem [{\citenamefont {Wilde}(2013)}]{wilde2013quantum}%
	\BibitemOpen
	\bibfield  {author} {\bibinfo {author} {\bibfnamefont {Mark~M}\ \bibnamefont
			{Wilde}},\ }\href@noop {} {\emph {\bibinfo {title} {Quantum Information
				Theory}}}\ (\bibinfo  {publisher} {Cambridge University Press},\ \bibinfo
	{year} {2013})\BibitemShut {NoStop}%
	\bibitem [{\citenamefont {Xu}\ \emph {et~al.}(2020)\citenamefont {Xu},
		\citenamefont {Ma}, \citenamefont {Zhang}, \citenamefont {Lo},\ and\
		\citenamefont {Pan}}]{xu20}%
	\BibitemOpen
	\bibfield  {author} {\bibinfo {author} {\bibfnamefont {Feihu}\ \bibnamefont
			{Xu}}, \bibinfo {author} {\bibfnamefont {Xiongfeng}\ \bibnamefont {Ma}},
		\bibinfo {author} {\bibfnamefont {Qiang}\ \bibnamefont {Zhang}}, \bibinfo
		{author} {\bibfnamefont {Hoi-Kwong}\ \bibnamefont {Lo}}, \ and\ \bibinfo
		{author} {\bibfnamefont {Jian-Wei}\ \bibnamefont {Pan}},\ }\bibfield  {title}
	{\enquote {\bibinfo {title} {Secure quantum key distribution with realistic
				devices},}\ }\href {https://link.aps.org/doi/10.1103/RevModPhys.92.025002}
	{\bibfield  {journal} {\bibinfo  {journal} {Rev. Mod. Phys.}\ }\textbf
		{\bibinfo {volume} {92}},\ \bibinfo {pages} {025002} (\bibinfo {year}
		{2020})}\BibitemShut {NoStop}%
	\bibitem [{\citenamefont {Cuomo}\ \emph {et~al.}(2020)\citenamefont {Cuomo},
		\citenamefont {Caleffi},\ and\ \citenamefont {Cacciapuoti}}]{cuomo20}%
	\BibitemOpen
	\bibfield  {author} {\bibinfo {author} {\bibfnamefont {Daniele}\ \bibnamefont
			{Cuomo}}, \bibinfo {author} {\bibfnamefont {Marcello}\ \bibnamefont
			{Caleffi}}, \ and\ \bibinfo {author} {\bibfnamefont {Angela~Sara}\
			\bibnamefont {Cacciapuoti}},\ }\bibfield  {title} {\enquote {\bibinfo {title}
			{Towards a distributed quantum computing ecosystem},}\ }\href
	{https://ietresearch.onlinelibrary.wiley.com/doi/full/10.1049/iet-qtc.2020.0002}
	{\bibfield  {journal} {\bibinfo  {journal} {IET Quantum Communication}\
		}\textbf {\bibinfo {volume} {1}},\ \bibinfo {pages} {3--8} (\bibinfo {year}
		{2020})}\BibitemShut {NoStop}%
	\bibitem [{\citenamefont {Wehner}\ \emph {et~al.}(2018)\citenamefont {Wehner},
		\citenamefont {Elkouss},\ and\ \citenamefont {Hanson}}]{wehner18}%
	\BibitemOpen
	\bibfield  {author} {\bibinfo {author} {\bibfnamefont {Stephanie}\
			\bibnamefont {Wehner}}, \bibinfo {author} {\bibfnamefont {David}\
			\bibnamefont {Elkouss}}, \ and\ \bibinfo {author} {\bibfnamefont {Ronald}\
			\bibnamefont {Hanson}},\ }\bibfield  {title} {\enquote {\bibinfo {title}
			{Quantum internet: A vision for the road ahead},}\ }\href
	{https://science.sciencemag.org/content/362/6412/eaam9288} {\bibfield
		{journal} {\bibinfo  {journal} {Science}\ }\textbf {\bibinfo {volume} {362}}
		(\bibinfo {year} {2018})}\BibitemShut {NoStop}%
	\bibitem [{\citenamefont {Steinlechner}\ \emph {et~al.}(2012)\citenamefont
		{Steinlechner}, \citenamefont {Trojek}, \citenamefont {Jofre}, \citenamefont
		{Weier}, \citenamefont {Perez}, \citenamefont {Jennewein}, \citenamefont
		{Ursin}, \citenamefont {Rarity}, \citenamefont {Mitchell}, \citenamefont
		{Torres}, \citenamefont {Weinfurter},\ and\ \citenamefont
		{Pruneri}}]{steinlechner12}%
	\BibitemOpen
	\bibfield  {author} {\bibinfo {author} {\bibfnamefont {Fabian}\ \bibnamefont
			{Steinlechner}}, \bibinfo {author} {\bibfnamefont {Pavel}\ \bibnamefont
			{Trojek}}, \bibinfo {author} {\bibfnamefont {Marc}\ \bibnamefont {Jofre}},
		\bibinfo {author} {\bibfnamefont {Henning}\ \bibnamefont {Weier}}, \bibinfo
		{author} {\bibfnamefont {Daniel}\ \bibnamefont {Perez}}, \bibinfo {author}
		{\bibfnamefont {Thomas}\ \bibnamefont {Jennewein}}, \bibinfo {author}
		{\bibfnamefont {Rupert}\ \bibnamefont {Ursin}}, \bibinfo {author}
		{\bibfnamefont {John}\ \bibnamefont {Rarity}}, \bibinfo {author}
		{\bibfnamefont {Morgan~W.}\ \bibnamefont {Mitchell}}, \bibinfo {author}
		{\bibfnamefont {Juan~P.}\ \bibnamefont {Torres}}, \bibinfo {author}
		{\bibfnamefont {Harald}\ \bibnamefont {Weinfurter}}, \ and\ \bibinfo {author}
		{\bibfnamefont {Valerio}\ \bibnamefont {Pruneri}},\ }\bibfield  {title}
	{\enquote {\bibinfo {title} {A high-brightness source of
				polarization-entangled photons optimized for applications in free space},}\
	}\href {\doibase 10.1364/OE.20.009640} {\bibfield  {journal} {\bibinfo
			{journal} {Opt. Express}\ }\textbf {\bibinfo {volume} {20}},\ \bibinfo
		{pages} {9640--9649} (\bibinfo {year} {2012})}\BibitemShut {NoStop}%
	\bibitem [{\citenamefont {Chen}\ \emph {et~al.}(2018)\citenamefont {Chen},
		\citenamefont {Ecker}, \citenamefont {Wengerowsky}, \citenamefont {Bulla},
		\citenamefont {Joshi}, \citenamefont {Steinlechner},\ and\ \citenamefont
		{Ursin}}]{chen18}%
	\BibitemOpen
	\bibfield  {author} {\bibinfo {author} {\bibfnamefont {Yuanyuan}\
			\bibnamefont {Chen}}, \bibinfo {author} {\bibfnamefont {Sebastian}\
			\bibnamefont {Ecker}}, \bibinfo {author} {\bibfnamefont {S\"oren}\
			\bibnamefont {Wengerowsky}}, \bibinfo {author} {\bibfnamefont {Lukas}\
			\bibnamefont {Bulla}}, \bibinfo {author} {\bibfnamefont {Siddarth~Koduru}\
			\bibnamefont {Joshi}}, \bibinfo {author} {\bibfnamefont {Fabian}\
			\bibnamefont {Steinlechner}}, \ and\ \bibinfo {author} {\bibfnamefont
			{Rupert}\ \bibnamefont {Ursin}},\ }\bibfield  {title} {\enquote {\bibinfo
			{title} {Polarization entanglement by time-reversed hong-ou-mandel
				interference},}\ }\href {\doibase 10.1103/PhysRevLett.121.200502} {\bibfield
		{journal} {\bibinfo  {journal} {Phys. Rev. Lett.}\ }\textbf {\bibinfo
			{volume} {121}},\ \bibinfo {pages} {200502} (\bibinfo {year}
		{2018})}\BibitemShut {NoStop}%
	\bibitem [{\citenamefont {Ecker}\ \emph {et~al.}(2021)\citenamefont {Ecker},
		\citenamefont {Liu}, \citenamefont {Handsteiner}, \citenamefont {Fink},
		\citenamefont {Rauch}, \citenamefont {Steinlechner}, \citenamefont {Scheidl},
		\citenamefont {Zeilinger},\ and\ \citenamefont {Ursin}}]{ecker2021}%
	\BibitemOpen
	\bibfield  {author} {\bibinfo {author} {\bibfnamefont {Sebastian}\
			\bibnamefont {Ecker}}, \bibinfo {author} {\bibfnamefont {Bo}~\bibnamefont
			{Liu}}, \bibinfo {author} {\bibfnamefont {Johannes}\ \bibnamefont
			{Handsteiner}}, \bibinfo {author} {\bibfnamefont {Matthias}\ \bibnamefont
			{Fink}}, \bibinfo {author} {\bibfnamefont {Dominik}\ \bibnamefont {Rauch}},
		\bibinfo {author} {\bibfnamefont {Fabian}\ \bibnamefont {Steinlechner}},
		\bibinfo {author} {\bibfnamefont {Thomas}\ \bibnamefont {Scheidl}}, \bibinfo
		{author} {\bibfnamefont {Anton}\ \bibnamefont {Zeilinger}}, \ and\ \bibinfo
		{author} {\bibfnamefont {Rupert}\ \bibnamefont {Ursin}},\ }\bibfield  {title}
	{\enquote {\bibinfo {title} {Strategies for achieving high key rates in
				satellite-based qkd},}\ }\href
	{https://www.nature.com/articles/s41534-020-00335-5} {\bibfield  {journal}
		{\bibinfo  {journal} {npj Quantum Information}\ }\textbf {\bibinfo {volume}
			{7}},\ \bibinfo {pages} {1--7} (\bibinfo {year} {2021})}\BibitemShut
	{NoStop}%
	\bibitem [{\citenamefont {Kues}\ \emph {et~al.}(2017)\citenamefont {Kues},
		\citenamefont {Reimer}, \citenamefont {Roztocki}, \citenamefont {Cort{\'e}s},
		\citenamefont {Sciara}, \citenamefont {Wetzel}, \citenamefont {Zhang},
		\citenamefont {Cino}, \citenamefont {Chu}, \citenamefont {Little} \emph
		{et~al.}}]{kues2017}%
	\BibitemOpen
	\bibfield  {author} {\bibinfo {author} {\bibfnamefont {Michael}\ \bibnamefont
			{Kues}}, \bibinfo {author} {\bibfnamefont {Christian}\ \bibnamefont
			{Reimer}}, \bibinfo {author} {\bibfnamefont {Piotr}\ \bibnamefont
			{Roztocki}}, \bibinfo {author} {\bibfnamefont {Luis~Romero}\ \bibnamefont
			{Cort{\'e}s}}, \bibinfo {author} {\bibfnamefont {Stefania}\ \bibnamefont
			{Sciara}}, \bibinfo {author} {\bibfnamefont {Benjamin}\ \bibnamefont
			{Wetzel}}, \bibinfo {author} {\bibfnamefont {Yanbing}\ \bibnamefont {Zhang}},
		\bibinfo {author} {\bibfnamefont {Alfonso}\ \bibnamefont {Cino}}, \bibinfo
		{author} {\bibfnamefont {Sai~T}\ \bibnamefont {Chu}}, \bibinfo {author}
		{\bibfnamefont {Brent~E}\ \bibnamefont {Little}},  \emph {et~al.},\
	}\bibfield  {title} {\enquote {\bibinfo {title} {On-chip generation of
				high-dimensional entangled quantum states and their coherent control},}\
	}\href {https://www.nature.com/articles/nature22986} {\bibfield  {journal}
		{\bibinfo  {journal} {Nature}\ }\textbf {\bibinfo {volume} {546}},\ \bibinfo
		{pages} {622--626} (\bibinfo {year} {2017})}\BibitemShut {NoStop}%
	\bibitem [{\citenamefont {Yin}\ \emph {et~al.}(2017)\citenamefont {Yin},
		\citenamefont {Cao}, \citenamefont {Li}, \citenamefont {Liao}, \citenamefont
		{Zhang}, \citenamefont {Ren}, \citenamefont {Cai}, \citenamefont {Liu},
		\citenamefont {Li}, \citenamefont {Dai}, \citenamefont {Li}, \citenamefont
		{Lu}, \citenamefont {Gong}, \citenamefont {Xu}, \citenamefont {Li},
		\citenamefont {Li}, \citenamefont {Yin}, \citenamefont {Jiang}, \citenamefont
		{Li}, \citenamefont {Jia}, \citenamefont {Ren}, \citenamefont {He},
		\citenamefont {Zhou}, \citenamefont {Zhang}, \citenamefont {Wang},
		\citenamefont {Chang}, \citenamefont {Zhu}, \citenamefont {Liu},
		\citenamefont {Chen}, \citenamefont {Lu}, \citenamefont {Shu}, \citenamefont
		{Peng}, \citenamefont {Wang},\ and\ \citenamefont {Pan}}]{yin17}%
	\BibitemOpen
	\bibfield  {author} {\bibinfo {author} {\bibfnamefont {Juan}\ \bibnamefont
			{Yin}}, \bibinfo {author} {\bibfnamefont {Yuan}\ \bibnamefont {Cao}},
		\bibinfo {author} {\bibfnamefont {Yu-Huai}\ \bibnamefont {Li}}, \bibinfo
		{author} {\bibfnamefont {Sheng-Kai}\ \bibnamefont {Liao}}, \bibinfo {author}
		{\bibfnamefont {Liang}\ \bibnamefont {Zhang}}, \bibinfo {author}
		{\bibfnamefont {Ji-Gang}\ \bibnamefont {Ren}}, \bibinfo {author}
		{\bibfnamefont {Wen-Qi}\ \bibnamefont {Cai}}, \bibinfo {author}
		{\bibfnamefont {Wei-Yue}\ \bibnamefont {Liu}}, \bibinfo {author}
		{\bibfnamefont {Bo}~\bibnamefont {Li}}, \bibinfo {author} {\bibfnamefont
			{Hui}\ \bibnamefont {Dai}}, \bibinfo {author} {\bibfnamefont {Guang-Bing}\
			\bibnamefont {Li}}, \bibinfo {author} {\bibfnamefont {Qi-Ming}\ \bibnamefont
			{Lu}}, \bibinfo {author} {\bibfnamefont {Yun-Hong}\ \bibnamefont {Gong}},
		\bibinfo {author} {\bibfnamefont {Yu}~\bibnamefont {Xu}}, \bibinfo {author}
		{\bibfnamefont {Shuang-Lin}\ \bibnamefont {Li}}, \bibinfo {author}
		{\bibfnamefont {Feng-Zhi}\ \bibnamefont {Li}}, \bibinfo {author}
		{\bibfnamefont {Ya-Yun}\ \bibnamefont {Yin}}, \bibinfo {author}
		{\bibfnamefont {Zi-Qing}\ \bibnamefont {Jiang}}, \bibinfo {author}
		{\bibfnamefont {Ming}\ \bibnamefont {Li}}, \bibinfo {author} {\bibfnamefont
			{Jian-Jun}\ \bibnamefont {Jia}}, \bibinfo {author} {\bibfnamefont
			{Ge}~\bibnamefont {Ren}}, \bibinfo {author} {\bibfnamefont {Dong}\
			\bibnamefont {He}}, \bibinfo {author} {\bibfnamefont {Yi-Lin}\ \bibnamefont
			{Zhou}}, \bibinfo {author} {\bibfnamefont {Xiao-Xiang}\ \bibnamefont
			{Zhang}}, \bibinfo {author} {\bibfnamefont {Na}~\bibnamefont {Wang}},
		\bibinfo {author} {\bibfnamefont {Xiang}\ \bibnamefont {Chang}}, \bibinfo
		{author} {\bibfnamefont {Zhen-Cai}\ \bibnamefont {Zhu}}, \bibinfo {author}
		{\bibfnamefont {Nai-Le}\ \bibnamefont {Liu}}, \bibinfo {author}
		{\bibfnamefont {Yu-Ao}\ \bibnamefont {Chen}}, \bibinfo {author}
		{\bibfnamefont {Chao-Yang}\ \bibnamefont {Lu}}, \bibinfo {author}
		{\bibfnamefont {Rong}\ \bibnamefont {Shu}}, \bibinfo {author} {\bibfnamefont
			{Cheng-Zhi}\ \bibnamefont {Peng}}, \bibinfo {author} {\bibfnamefont
			{Jian-Yu}\ \bibnamefont {Wang}}, \ and\ \bibinfo {author} {\bibfnamefont
			{Jian-Wei}\ \bibnamefont {Pan}},\ }\bibfield  {title} {\enquote {\bibinfo
			{title} {Satellite-based entanglement distribution over 1200 kilometers},}\
	}\href {\doibase 10.1126/science.aan3211} {\bibfield  {journal} {\bibinfo
			{journal} {Science}\ }\textbf {\bibinfo {volume} {356}},\ \bibinfo {pages}
		{1140--1144} (\bibinfo {year} {2017})}\BibitemShut {NoStop}%
	\bibitem [{\citenamefont {Joshi}\ \emph {et~al.}(2020)\citenamefont {Joshi},
		\citenamefont {Aktas}, \citenamefont {Wengerowsky}, \citenamefont {Lon{\v
				c}ari{\'c}}, \citenamefont {Neumann}, \citenamefont {Liu}, \citenamefont
		{Scheidl}, \citenamefont {Lorenzo}, \citenamefont {Samec}, \citenamefont
		{Kling}, \citenamefont {Qiu}, \citenamefont {Razavi}, \citenamefont {Stip{\v
				c}evi{\'c}}, \citenamefont {Rarity},\ and\ \citenamefont {Ursin}}]{joshi20}%
	\BibitemOpen
	\bibfield  {author} {\bibinfo {author} {\bibfnamefont {Siddarth~Koduru}\
			\bibnamefont {Joshi}}, \bibinfo {author} {\bibfnamefont {Djeylan}\
			\bibnamefont {Aktas}}, \bibinfo {author} {\bibfnamefont {S{\"o}ren}\
			\bibnamefont {Wengerowsky}}, \bibinfo {author} {\bibfnamefont {Martin}\
			\bibnamefont {Lon{\v c}ari{\'c}}}, \bibinfo {author} {\bibfnamefont
			{Sebastian~Philipp}\ \bibnamefont {Neumann}}, \bibinfo {author}
		{\bibfnamefont {Bo}~\bibnamefont {Liu}}, \bibinfo {author} {\bibfnamefont
			{Thomas}\ \bibnamefont {Scheidl}}, \bibinfo {author} {\bibfnamefont
			{Guillermo~Curr{\'a}s}\ \bibnamefont {Lorenzo}}, \bibinfo {author}
		{\bibfnamefont {{\v Z}eljko}\ \bibnamefont {Samec}}, \bibinfo {author}
		{\bibfnamefont {Laurent}\ \bibnamefont {Kling}}, \bibinfo {author}
		{\bibfnamefont {Alex}\ \bibnamefont {Qiu}}, \bibinfo {author} {\bibfnamefont
			{Mohsen}\ \bibnamefont {Razavi}}, \bibinfo {author} {\bibfnamefont {Mario}\
			\bibnamefont {Stip{\v c}evi{\'c}}}, \bibinfo {author} {\bibfnamefont
			{John~G.}\ \bibnamefont {Rarity}}, \ and\ \bibinfo {author} {\bibfnamefont
			{Rupert}\ \bibnamefont {Ursin}},\ }\bibfield  {title} {\enquote {\bibinfo
			{title} {A trusted node{\textendash}free eight-user metropolitan quantum
				communication network},}\ }\href {\doibase 10.1126/sciadv.aba0959} {\bibfield
		{journal} {\bibinfo  {journal} {Science Advances}\ }\textbf {\bibinfo
			{volume} {6}} (\bibinfo {year} {2020}),\ 10.1126/sciadv.aba0959}\BibitemShut
	{NoStop}%
	\bibitem [{\citenamefont {Valivarthi}\ \emph {et~al.}(2016)\citenamefont
		{Valivarthi}, \citenamefont {Zhou}, \citenamefont {Aguilar}, \citenamefont
		{Verma}, \citenamefont {Marsili}, \citenamefont {Shaw}, \citenamefont {Nam},
		\citenamefont {Oblak}, \citenamefont {Tittel} \emph {et~al.}}]{valivarthi16}%
	\BibitemOpen
	\bibfield  {author} {\bibinfo {author} {\bibfnamefont {Raju}\ \bibnamefont
			{Valivarthi}}, \bibinfo {author} {\bibfnamefont {Qiang}\ \bibnamefont
			{Zhou}}, \bibinfo {author} {\bibfnamefont {Gabriel~H}\ \bibnamefont
			{Aguilar}}, \bibinfo {author} {\bibfnamefont {Varun~B}\ \bibnamefont
			{Verma}}, \bibinfo {author} {\bibfnamefont {Francesco}\ \bibnamefont
			{Marsili}}, \bibinfo {author} {\bibfnamefont {Matthew~D}\ \bibnamefont
			{Shaw}}, \bibinfo {author} {\bibfnamefont {Sae~Woo}\ \bibnamefont {Nam}},
		\bibinfo {author} {\bibfnamefont {Daniel}\ \bibnamefont {Oblak}}, \bibinfo
		{author} {\bibfnamefont {Wolfgang}\ \bibnamefont {Tittel}},  \emph {et~al.},\
	}\bibfield  {title} {\enquote {\bibinfo {title} {Quantum teleportation across
				a metropolitan fibre network},}\ }\href
	{https://www.nature.com/articles/nphoton.2016.180} {\bibfield  {journal}
		{\bibinfo  {journal} {Nature Photonics}\ }\textbf {\bibinfo {volume} {10}},\
		\bibinfo {pages} {676--680} (\bibinfo {year} {2016})}\BibitemShut {NoStop}%
	\bibitem [{\citenamefont {Wengerowsky}\ \emph {et~al.}(2019)\citenamefont
		{Wengerowsky}, \citenamefont {Joshi}, \citenamefont {Steinlechner},
		\citenamefont {Zichi}, \citenamefont {Dobrovolskiy}, \citenamefont {van~der
			Molen}, \citenamefont {Los}, \citenamefont {Zwiller}, \citenamefont
		{Versteegh}, \citenamefont {Mura}, \citenamefont {Calonico}, \citenamefont
		{Inguscio}, \citenamefont {H{\"u}bel}, \citenamefont {Bo}, \citenamefont
		{Scheidl}, \citenamefont {Zeilinger}, \citenamefont {Xuereb},\ and\
		\citenamefont {Ursin}}]{Wengerowsky19}%
	\BibitemOpen
	\bibfield  {author} {\bibinfo {author} {\bibfnamefont {S{\"o}ren}\
			\bibnamefont {Wengerowsky}}, \bibinfo {author} {\bibfnamefont
			{Siddarth~Koduru}\ \bibnamefont {Joshi}}, \bibinfo {author} {\bibfnamefont
			{Fabian}\ \bibnamefont {Steinlechner}}, \bibinfo {author} {\bibfnamefont
			{Julien~R.}\ \bibnamefont {Zichi}}, \bibinfo {author} {\bibfnamefont
			{Sergiy~M.}\ \bibnamefont {Dobrovolskiy}}, \bibinfo {author} {\bibfnamefont
			{Ren{\'e}}\ \bibnamefont {van~der Molen}}, \bibinfo {author} {\bibfnamefont
			{Johannes W.~N.}\ \bibnamefont {Los}}, \bibinfo {author} {\bibfnamefont
			{Val}\ \bibnamefont {Zwiller}}, \bibinfo {author} {\bibfnamefont {Marijn
				A.~M.}\ \bibnamefont {Versteegh}}, \bibinfo {author} {\bibfnamefont
			{Alberto}\ \bibnamefont {Mura}}, \bibinfo {author} {\bibfnamefont {Davide}\
			\bibnamefont {Calonico}}, \bibinfo {author} {\bibfnamefont {Massimo}\
			\bibnamefont {Inguscio}}, \bibinfo {author} {\bibfnamefont {Hannes}\
			\bibnamefont {H{\"u}bel}}, \bibinfo {author} {\bibfnamefont {Liu}\
			\bibnamefont {Bo}}, \bibinfo {author} {\bibfnamefont {Thomas}\ \bibnamefont
			{Scheidl}}, \bibinfo {author} {\bibfnamefont {Anton}\ \bibnamefont
			{Zeilinger}}, \bibinfo {author} {\bibfnamefont {Andr{\'e}}\ \bibnamefont
			{Xuereb}}, \ and\ \bibinfo {author} {\bibfnamefont {Rupert}\ \bibnamefont
			{Ursin}},\ }\bibfield  {title} {\enquote {\bibinfo {title} {Entanglement
				distribution over a 96-km-long submarine optical fiber},}\ }\href {\doibase
		10.1073/pnas.1818752116} {\bibfield  {journal} {\bibinfo  {journal}
			{Proceedings of the National Academy of Sciences}\ }\textbf {\bibinfo
			{volume} {116}},\ \bibinfo {pages} {6684--6688} (\bibinfo {year}
		{2019})}\BibitemShut {NoStop}%
	\bibitem [{\citenamefont {Schlosshauer}(2019)}]{schlosshauer19}%
	\BibitemOpen
	\bibfield  {author} {\bibinfo {author} {\bibfnamefont {Maximilian}\
			\bibnamefont {Schlosshauer}},\ }\bibfield  {title} {\enquote {\bibinfo
			{title} {Quantum decoherence},}\ }\href {\doibase
		https://doi.org/10.1016/j.physrep.2019.10.001} {\bibfield  {journal}
		{\bibinfo  {journal} {Physics Reports}\ }\textbf {\bibinfo {volume} {831}},\
		\bibinfo {pages} {1 -- 57} (\bibinfo {year} {2019})}\BibitemShut {NoStop}%
	\bibitem [{\citenamefont {Bennett}\ \emph
		{et~al.}(1996{\natexlab{a}})\citenamefont {Bennett}, \citenamefont
		{Brassard}, \citenamefont {Popescu}, \citenamefont {Schumacher},
		\citenamefont {Smolin},\ and\ \citenamefont {Wootters}}]{bennett96}%
	\BibitemOpen
	\bibfield  {author} {\bibinfo {author} {\bibfnamefont {Charles~H.}\
			\bibnamefont {Bennett}}, \bibinfo {author} {\bibfnamefont {Gilles}\
			\bibnamefont {Brassard}}, \bibinfo {author} {\bibfnamefont {Sandu}\
			\bibnamefont {Popescu}}, \bibinfo {author} {\bibfnamefont {Benjamin}\
			\bibnamefont {Schumacher}}, \bibinfo {author} {\bibfnamefont {John~A.}\
			\bibnamefont {Smolin}}, \ and\ \bibinfo {author} {\bibfnamefont {William~K.}\
			\bibnamefont {Wootters}},\ }\bibfield  {title} {\enquote {\bibinfo {title}
			{Purification of noisy entanglement and faithful teleportation via noisy
				channels},}\ }\href {\doibase 10.1103/PhysRevLett.76.722} {\bibfield
		{journal} {\bibinfo  {journal} {Phys. Rev. Lett.}\ }\textbf {\bibinfo
			{volume} {76}},\ \bibinfo {pages} {722--725} (\bibinfo {year}
		{1996}{\natexlab{a}})}\BibitemShut {NoStop}%
	\bibitem [{\citenamefont {Deutsch}\ \emph {et~al.}(1996)\citenamefont
		{Deutsch}, \citenamefont {Ekert}, \citenamefont {Jozsa}, \citenamefont
		{Macchiavello}, \citenamefont {Popescu},\ and\ \citenamefont
		{Sanpera}}]{deutsch96}%
	\BibitemOpen
	\bibfield  {author} {\bibinfo {author} {\bibfnamefont {David}\ \bibnamefont
			{Deutsch}}, \bibinfo {author} {\bibfnamefont {Artur}\ \bibnamefont {Ekert}},
		\bibinfo {author} {\bibfnamefont {Richard}\ \bibnamefont {Jozsa}}, \bibinfo
		{author} {\bibfnamefont {Chiara}\ \bibnamefont {Macchiavello}}, \bibinfo
		{author} {\bibfnamefont {Sandu}\ \bibnamefont {Popescu}}, \ and\ \bibinfo
		{author} {\bibfnamefont {Anna}\ \bibnamefont {Sanpera}},\ }\bibfield  {title}
	{\enquote {\bibinfo {title} {Quantum privacy amplification and the security
				of quantum cryptography over noisy channels},}\ }\href {\doibase
		10.1103/PhysRevLett.77.2818} {\bibfield  {journal} {\bibinfo  {journal}
			{Phys. Rev. Lett.}\ }\textbf {\bibinfo {volume} {77}},\ \bibinfo {pages}
		{2818--2821} (\bibinfo {year} {1996})}\BibitemShut {NoStop}%
	\bibitem [{\citenamefont {Briegel}\ \emph {et~al.}(1998)\citenamefont
		{Briegel}, \citenamefont {D\"ur}, \citenamefont {Cirac},\ and\ \citenamefont
		{Zoller}}]{briegel98}%
	\BibitemOpen
	\bibfield  {author} {\bibinfo {author} {\bibfnamefont {H.-J.}\ \bibnamefont
			{Briegel}}, \bibinfo {author} {\bibfnamefont {W.}~\bibnamefont {D\"ur}},
		\bibinfo {author} {\bibfnamefont {J.~I.}\ \bibnamefont {Cirac}}, \ and\
		\bibinfo {author} {\bibfnamefont {P.}~\bibnamefont {Zoller}},\ }\bibfield
	{title} {\enquote {\bibinfo {title} {Quantum repeaters: The role of imperfect
				local operations in quantum communication},}\ }\href {\doibase
		10.1103/PhysRevLett.81.5932} {\bibfield  {journal} {\bibinfo  {journal}
			{Phys. Rev. Lett.}\ }\textbf {\bibinfo {volume} {81}},\ \bibinfo {pages}
		{5932--5935} (\bibinfo {year} {1998})}\BibitemShut {NoStop}%
	\bibitem [{\citenamefont {D\"ur}\ \emph {et~al.}(1999)\citenamefont {D\"ur},
		\citenamefont {Briegel}, \citenamefont {Cirac},\ and\ \citenamefont
		{Zoller}}]{duer99}%
	\BibitemOpen
	\bibfield  {author} {\bibinfo {author} {\bibfnamefont {W.}~\bibnamefont
			{D\"ur}}, \bibinfo {author} {\bibfnamefont {H.-J.}\ \bibnamefont {Briegel}},
		\bibinfo {author} {\bibfnamefont {J.~I.}\ \bibnamefont {Cirac}}, \ and\
		\bibinfo {author} {\bibfnamefont {P.}~\bibnamefont {Zoller}},\ }\bibfield
	{title} {\enquote {\bibinfo {title} {Quantum repeaters based on entanglement
				purification},}\ }\href {\doibase 10.1103/PhysRevA.59.169} {\bibfield
		{journal} {\bibinfo  {journal} {Phys. Rev. A}\ }\textbf {\bibinfo {volume}
			{59}},\ \bibinfo {pages} {169--181} (\bibinfo {year} {1999})}\BibitemShut
	{NoStop}%
	\bibitem [{\citenamefont {Chen}\ \emph {et~al.}(2017)\citenamefont {Chen},
		\citenamefont {Yong}, \citenamefont {Xu}, \citenamefont {Yao}, \citenamefont
		{Xiang}, \citenamefont {Li}, \citenamefont {Liu}, \citenamefont {Lu},
		\citenamefont {Liu}, \citenamefont {Li} \emph {et~al.}}]{chen17}%
	\BibitemOpen
	\bibfield  {author} {\bibinfo {author} {\bibfnamefont {Luo-Kan}\ \bibnamefont
			{Chen}}, \bibinfo {author} {\bibfnamefont {Hai-Lin}\ \bibnamefont {Yong}},
		\bibinfo {author} {\bibfnamefont {Ping}\ \bibnamefont {Xu}}, \bibinfo
		{author} {\bibfnamefont {Xing-Can}\ \bibnamefont {Yao}}, \bibinfo {author}
		{\bibfnamefont {Tong}\ \bibnamefont {Xiang}}, \bibinfo {author}
		{\bibfnamefont {Zheng-Da}\ \bibnamefont {Li}}, \bibinfo {author}
		{\bibfnamefont {Chang}\ \bibnamefont {Liu}}, \bibinfo {author} {\bibfnamefont
			{He}~\bibnamefont {Lu}}, \bibinfo {author} {\bibfnamefont {Nai-Le}\
			\bibnamefont {Liu}}, \bibinfo {author} {\bibfnamefont {Li}~\bibnamefont
			{Li}},  \emph {et~al.},\ }\bibfield  {title} {\enquote {\bibinfo {title}
			{Experimental nested purification for a linear optical quantum repeater},}\
	}\href {https://www.nature.com/articles/s41566-017-0010-6} {\bibfield
		{journal} {\bibinfo  {journal} {Nature Photonics}\ }\textbf {\bibinfo
			{volume} {11}},\ \bibinfo {pages} {695--699} (\bibinfo {year}
		{2017})}\BibitemShut {NoStop}%
	\bibitem [{\citenamefont {O'Brien}\ \emph {et~al.}(2003)\citenamefont
		{O'Brien}, \citenamefont {Pryde}, \citenamefont {White}, \citenamefont
		{Ralph},\ and\ \citenamefont {Branning}}]{obrien03}%
	\BibitemOpen
	\bibfield  {author} {\bibinfo {author} {\bibfnamefont {Jeremy~L}\
			\bibnamefont {O'Brien}}, \bibinfo {author} {\bibfnamefont {Geoffrey~J}\
			\bibnamefont {Pryde}}, \bibinfo {author} {\bibfnamefont {Andrew~G}\
			\bibnamefont {White}}, \bibinfo {author} {\bibfnamefont {Timothy~C}\
			\bibnamefont {Ralph}}, \ and\ \bibinfo {author} {\bibfnamefont {David}\
			\bibnamefont {Branning}},\ }\bibfield  {title} {\enquote {\bibinfo {title}
			{Demonstration of an all-optical quantum controlled-not gate},}\ }\href
	{https://www.nature.com/articles/nature02054} {\bibfield  {journal} {\bibinfo
			{journal} {Nature}\ }\textbf {\bibinfo {volume} {426}},\ \bibinfo {pages}
		{264--267} (\bibinfo {year} {2003})}\BibitemShut {NoStop}%
	\bibitem [{\citenamefont {Pittman}\ \emph {et~al.}(2003)\citenamefont
		{Pittman}, \citenamefont {Fitch}, \citenamefont {Jacobs},\ and\ \citenamefont
		{Franson}}]{pittman03}%
	\BibitemOpen
	\bibfield  {author} {\bibinfo {author} {\bibfnamefont {T.~B.}\ \bibnamefont
			{Pittman}}, \bibinfo {author} {\bibfnamefont {M.~J.}\ \bibnamefont {Fitch}},
		\bibinfo {author} {\bibfnamefont {B.~C}\ \bibnamefont {Jacobs}}, \ and\
		\bibinfo {author} {\bibfnamefont {J.~D.}\ \bibnamefont {Franson}},\
	}\bibfield  {title} {\enquote {\bibinfo {title} {Experimental controlled-not
				logic gate for single photons in the coincidence basis},}\ }\href {\doibase
		10.1103/PhysRevA.68.032316} {\bibfield  {journal} {\bibinfo  {journal} {Phys.
				Rev. A}\ }\textbf {\bibinfo {volume} {68}},\ \bibinfo {pages} {032316}
		(\bibinfo {year} {2003})}\BibitemShut {NoStop}%
	\bibitem [{\citenamefont {Gasparoni}\ \emph {et~al.}(2004)\citenamefont
		{Gasparoni}, \citenamefont {Pan}, \citenamefont {Walther}, \citenamefont
		{Rudolph},\ and\ \citenamefont {Zeilinger}}]{gasparoni04}%
	\BibitemOpen
	\bibfield  {author} {\bibinfo {author} {\bibfnamefont {Sara}\ \bibnamefont
			{Gasparoni}}, \bibinfo {author} {\bibfnamefont {Jian-Wei}\ \bibnamefont
			{Pan}}, \bibinfo {author} {\bibfnamefont {Philip}\ \bibnamefont {Walther}},
		\bibinfo {author} {\bibfnamefont {Terry}\ \bibnamefont {Rudolph}}, \ and\
		\bibinfo {author} {\bibfnamefont {Anton}\ \bibnamefont {Zeilinger}},\
	}\bibfield  {title} {\enquote {\bibinfo {title} {Realization of a photonic
				controlled-not gate sufficient for quantum computation},}\ }\href {\doibase
		10.1103/PhysRevLett.93.020504} {\bibfield  {journal} {\bibinfo  {journal}
			{Phys. Rev. Lett.}\ }\textbf {\bibinfo {volume} {93}},\ \bibinfo {pages}
		{020504} (\bibinfo {year} {2004})}\BibitemShut {NoStop}%
	\bibitem [{\citenamefont {Zhao}\ \emph {et~al.}(2005)\citenamefont {Zhao},
		\citenamefont {Zhang}, \citenamefont {Chen}, \citenamefont {Zhang},
		\citenamefont {Du}, \citenamefont {Yang},\ and\ \citenamefont
		{Pan}}]{zhao05}%
	\BibitemOpen
	\bibfield  {author} {\bibinfo {author} {\bibfnamefont {Zhi}\ \bibnamefont
			{Zhao}}, \bibinfo {author} {\bibfnamefont {An-Ning}\ \bibnamefont {Zhang}},
		\bibinfo {author} {\bibfnamefont {Yu-Ao}\ \bibnamefont {Chen}}, \bibinfo
		{author} {\bibfnamefont {Han}\ \bibnamefont {Zhang}}, \bibinfo {author}
		{\bibfnamefont {Jiang-Feng}\ \bibnamefont {Du}}, \bibinfo {author}
		{\bibfnamefont {Tao}\ \bibnamefont {Yang}}, \ and\ \bibinfo {author}
		{\bibfnamefont {Jian-Wei}\ \bibnamefont {Pan}},\ }\bibfield  {title}
	{\enquote {\bibinfo {title} {Experimental demonstration of a nondestructive
				controlled-not quantum gate for two independent photon qubits},}\ }\href
	{\doibase 10.1103/PhysRevLett.94.030501} {\bibfield  {journal} {\bibinfo
			{journal} {Phys. Rev. Lett.}\ }\textbf {\bibinfo {volume} {94}},\ \bibinfo
		{pages} {030501} (\bibinfo {year} {2005})}\BibitemShut {NoStop}%
	\bibitem [{\citenamefont {Pan}\ \emph {et~al.}(2003)\citenamefont {Pan},
		\citenamefont {Gasparoni}, \citenamefont {Ursin}, \citenamefont {Weihs},\
		and\ \citenamefont {Zeilinger}}]{pan03}%
	\BibitemOpen
	\bibfield  {author} {\bibinfo {author} {\bibfnamefont {Jian-Wei}\
			\bibnamefont {Pan}}, \bibinfo {author} {\bibfnamefont {Sara}\ \bibnamefont
			{Gasparoni}}, \bibinfo {author} {\bibfnamefont {Rupert}\ \bibnamefont
			{Ursin}}, \bibinfo {author} {\bibfnamefont {Gregor}\ \bibnamefont {Weihs}}, \
		and\ \bibinfo {author} {\bibfnamefont {Anton}\ \bibnamefont {Zeilinger}},\
	}\bibfield  {title} {\enquote {\bibinfo {title} {Experimental entanglement
				purification of arbitrary unknown states},}\ }\href
	{https://www.nature.com/articles/nature01623} {\bibfield  {journal} {\bibinfo
			{journal} {Nature}\ }\textbf {\bibinfo {volume} {423}},\ \bibinfo {pages}
		{417--422} (\bibinfo {year} {2003})}\BibitemShut {NoStop}%
	\bibitem [{\citenamefont {Yamamoto}\ \emph {et~al.}(2003)\citenamefont
		{Yamamoto}, \citenamefont {Koashi}, \citenamefont {{\"O}zdemir},\ and\
		\citenamefont {Imoto}}]{yamamoto03}%
	\BibitemOpen
	\bibfield  {author} {\bibinfo {author} {\bibfnamefont {Takashi}\ \bibnamefont
			{Yamamoto}}, \bibinfo {author} {\bibfnamefont {Masato}\ \bibnamefont
			{Koashi}}, \bibinfo {author} {\bibfnamefont {{\c{S}}ahin~Kaya}\ \bibnamefont
			{{\"O}zdemir}}, \ and\ \bibinfo {author} {\bibfnamefont {Nobuyuki}\
			\bibnamefont {Imoto}},\ }\bibfield  {title} {\enquote {\bibinfo {title}
			{Experimental extraction of an entangled photon pair from two identically
				decohered pairs},}\ }\href {https://www.nature.com/articles/nature01358}
	{\bibfield  {journal} {\bibinfo  {journal} {Nature}\ }\textbf {\bibinfo
			{volume} {421}},\ \bibinfo {pages} {343--346} (\bibinfo {year}
		{2003})}\BibitemShut {NoStop}%
	\bibitem [{\citenamefont {Walther}\ \emph {et~al.}(2005)\citenamefont
		{Walther}, \citenamefont {Resch}, \citenamefont {Brukner}, \citenamefont
		{Steinberg}, \citenamefont {Pan},\ and\ \citenamefont
		{Zeilinger}}]{walther05}%
	\BibitemOpen
	\bibfield  {author} {\bibinfo {author} {\bibfnamefont {P.}~\bibnamefont
			{Walther}}, \bibinfo {author} {\bibfnamefont {K.~J.}\ \bibnamefont {Resch}},
		\bibinfo {author} {\bibfnamefont {\ifmmode \check{C}\else~\v{C}\fi{}.}\
			\bibnamefont {Brukner}}, \bibinfo {author} {\bibfnamefont {A.~M.}\
			\bibnamefont {Steinberg}}, \bibinfo {author} {\bibfnamefont {J.-W.}\
			\bibnamefont {Pan}}, \ and\ \bibinfo {author} {\bibfnamefont
			{A.}~\bibnamefont {Zeilinger}},\ }\bibfield  {title} {\enquote {\bibinfo
			{title} {Quantum nonlocality obtained from local states by entanglement
				purification},}\ }\href {\doibase 10.1103/PhysRevLett.94.040504} {\bibfield
		{journal} {\bibinfo  {journal} {Phys. Rev. Lett.}\ }\textbf {\bibinfo
			{volume} {94}},\ \bibinfo {pages} {040504} (\bibinfo {year}
		{2005})}\BibitemShut {NoStop}%
	\bibitem [{\citenamefont {Simon}\ and\ \citenamefont {Pan}(2002)}]{simon02}%
	\BibitemOpen
	\bibfield  {author} {\bibinfo {author} {\bibfnamefont {Christoph}\
			\bibnamefont {Simon}}\ and\ \bibinfo {author} {\bibfnamefont {Jian-Wei}\
			\bibnamefont {Pan}},\ }\bibfield  {title} {\enquote {\bibinfo {title}
			{Polarization entanglement purification using spatial entanglement},}\ }\href
	{\doibase 10.1103/PhysRevLett.89.257901} {\bibfield  {journal} {\bibinfo
			{journal} {Phys. Rev. Lett.}\ }\textbf {\bibinfo {volume} {89}},\ \bibinfo
		{pages} {257901} (\bibinfo {year} {2002})}\BibitemShut {NoStop}%
	\bibitem [{\citenamefont {Barreiro}\ \emph {et~al.}(2005)\citenamefont
		{Barreiro}, \citenamefont {Langford}, \citenamefont {Peters},\ and\
		\citenamefont {Kwiat}}]{Barreiro05}%
	\BibitemOpen
	\bibfield  {author} {\bibinfo {author} {\bibfnamefont {Julio~T.}\
			\bibnamefont {Barreiro}}, \bibinfo {author} {\bibfnamefont {Nathan~K.}\
			\bibnamefont {Langford}}, \bibinfo {author} {\bibfnamefont {Nicholas~A.}\
			\bibnamefont {Peters}}, \ and\ \bibinfo {author} {\bibfnamefont {Paul~G.}\
			\bibnamefont {Kwiat}},\ }\bibfield  {title} {\enquote {\bibinfo {title}
			{Generation of hyperentangled photon pairs},}\ }\href {\doibase
		10.1103/PhysRevLett.95.260501} {\bibfield  {journal} {\bibinfo  {journal}
			{Phys. Rev. Lett.}\ }\textbf {\bibinfo {volume} {95}},\ \bibinfo {pages}
		{260501} (\bibinfo {year} {2005})}\BibitemShut {NoStop}%
	\bibitem [{\citenamefont {Barbieri}\ \emph {et~al.}(2005)\citenamefont
		{Barbieri}, \citenamefont {Cinelli}, \citenamefont {Mataloni},\ and\
		\citenamefont {De~Martini}}]{barbieri05}%
	\BibitemOpen
	\bibfield  {author} {\bibinfo {author} {\bibfnamefont {M.}~\bibnamefont
			{Barbieri}}, \bibinfo {author} {\bibfnamefont {C.}~\bibnamefont {Cinelli}},
		\bibinfo {author} {\bibfnamefont {P.}~\bibnamefont {Mataloni}}, \ and\
		\bibinfo {author} {\bibfnamefont {F.}~\bibnamefont {De~Martini}},\ }\bibfield
	{title} {\enquote {\bibinfo {title} {Polarization-momentum hyperentangled
				states: Realization and characterization},}\ }\href
	{https://link.aps.org/doi/10.1103/PhysRevA.72.052110} {\bibfield  {journal}
		{\bibinfo  {journal} {Phys. Rev. A}\ }\textbf {\bibinfo {volume} {72}},\
		\bibinfo {pages} {052110} (\bibinfo {year} {2005})}\BibitemShut {NoStop}%
	\bibitem [{\citenamefont {Li}\ and\ \citenamefont {Ghose}(2016)}]{Li16}%
	\BibitemOpen
	\bibfield  {author} {\bibinfo {author} {\bibfnamefont {Xi-Han}\ \bibnamefont
			{Li}}\ and\ \bibinfo {author} {\bibfnamefont {Shohini}\ \bibnamefont
			{Ghose}},\ }\bibfield  {title} {\enquote {\bibinfo {title} {Complete
				hyperentangled bell state analysis for polarization and time-bin
				hyperentanglement},}\ }\href
	{http://www.opticsexpress.org/abstract.cfm?URI=oe-24-16-18388} {\bibfield
		{journal} {\bibinfo  {journal} {Opt. Express}\ }\textbf {\bibinfo {volume}
			{24}},\ \bibinfo {pages} {18388--18398} (\bibinfo {year} {2016})}\BibitemShut
	{NoStop}%
	\bibitem [{\citenamefont {Fiorentino}\ and\ \citenamefont
		{Wong}(2004)}]{fiorentino04}%
	\BibitemOpen
	\bibfield  {author} {\bibinfo {author} {\bibfnamefont {Marco}\ \bibnamefont
			{Fiorentino}}\ and\ \bibinfo {author} {\bibfnamefont {Franco N.~C.}\
			\bibnamefont {Wong}},\ }\bibfield  {title} {\enquote {\bibinfo {title}
			{Deterministic controlled-not gate for single-photon two-qubit quantum
				logic},}\ }\href {\doibase 10.1103/PhysRevLett.93.070502} {\bibfield
		{journal} {\bibinfo  {journal} {Phys. Rev. Lett.}\ }\textbf {\bibinfo
			{volume} {93}},\ \bibinfo {pages} {070502} (\bibinfo {year}
		{2004})}\BibitemShut {NoStop}%
	\bibitem [{\citenamefont {Barreiro}\ \emph {et~al.}(2008)\citenamefont
		{Barreiro}, \citenamefont {Wei},\ and\ \citenamefont {Kwiat}}]{barreiro08}%
	\BibitemOpen
	\bibfield  {author} {\bibinfo {author} {\bibfnamefont {Julio~T}\ \bibnamefont
			{Barreiro}}, \bibinfo {author} {\bibfnamefont {Tzu-Chieh}\ \bibnamefont
			{Wei}}, \ and\ \bibinfo {author} {\bibfnamefont {Paul~G}\ \bibnamefont
			{Kwiat}},\ }\bibfield  {title} {\enquote {\bibinfo {title} {Beating the
				channel capacity limit for linear photonic superdense coding},}\ }\href
	{https://www.nature.com/articles/nphys919} {\bibfield  {journal} {\bibinfo
			{journal} {Nature physics}\ }\textbf {\bibinfo {volume} {4}},\ \bibinfo
		{pages} {282--286} (\bibinfo {year} {2008})}\BibitemShut {NoStop}%
	\bibitem [{\citenamefont {Hu}\ \emph {et~al.}(2021)\citenamefont {Hu},
		\citenamefont {Huang}, \citenamefont {Sheng}, \citenamefont {Zhou},
		\citenamefont {Liu}, \citenamefont {Guo}, \citenamefont {Zhang},
		\citenamefont {Xing}, \citenamefont {Huang}, \citenamefont {Li},\ and\
		\citenamefont {Guo}}]{hu21}%
	\BibitemOpen
	\bibfield  {author} {\bibinfo {author} {\bibfnamefont {Xiao-Min}\
			\bibnamefont {Hu}}, \bibinfo {author} {\bibfnamefont {Cen-Xiao}\ \bibnamefont
			{Huang}}, \bibinfo {author} {\bibfnamefont {Yu-Bo}\ \bibnamefont {Sheng}},
		\bibinfo {author} {\bibfnamefont {Lan}\ \bibnamefont {Zhou}}, \bibinfo
		{author} {\bibfnamefont {Bi-Heng}\ \bibnamefont {Liu}}, \bibinfo {author}
		{\bibfnamefont {Yu}~\bibnamefont {Guo}}, \bibinfo {author} {\bibfnamefont
			{Chao}\ \bibnamefont {Zhang}}, \bibinfo {author} {\bibfnamefont {Wen-Bo}\
			\bibnamefont {Xing}}, \bibinfo {author} {\bibfnamefont {Yun-Feng}\
			\bibnamefont {Huang}}, \bibinfo {author} {\bibfnamefont {Chuan-Feng}\
			\bibnamefont {Li}}, \ and\ \bibinfo {author} {\bibfnamefont {Guang-Can}\
			\bibnamefont {Guo}},\ }\bibfield  {title} {\enquote {\bibinfo {title}
			{Long-distance entanglement purification for quantum communication},}\ }\href
	{\doibase 10.1103/PhysRevLett.126.010503} {\bibfield  {journal} {\bibinfo
			{journal} {Phys. Rev. Lett.}\ }\textbf {\bibinfo {volume} {126}},\ \bibinfo
		{pages} {010503} (\bibinfo {year} {2021})}\BibitemShut {NoStop}%
	\bibitem [{\citenamefont {Kwiat}(1997)}]{Kwiat97}%
	\BibitemOpen
	\bibfield  {author} {\bibinfo {author} {\bibfnamefont {Paul~G.}\ \bibnamefont
			{Kwiat}},\ }\bibfield  {title} {\enquote {\bibinfo {title} {Hyper-entangled
				states},}\ }\href {\doibase 10.1080/09500349708231877} {\bibfield  {journal}
		{\bibinfo  {journal} {Journal of Modern Optics}\ }\textbf {\bibinfo {volume}
			{44}},\ \bibinfo {pages} {2173--2184} (\bibinfo {year} {1997})}\BibitemShut
	{NoStop}%
	\bibitem [{\citenamefont {Prilm\"uller}\ \emph {et~al.}(2018)\citenamefont
		{Prilm\"uller}, \citenamefont {Huber}, \citenamefont {M\"uller},
		\citenamefont {Michler}, \citenamefont {Weihs},\ and\ \citenamefont
		{Predojevi\ifmmode~\acute{c}\else \'{c}\fi{}}}]{prilm18}%
	\BibitemOpen
	\bibfield  {author} {\bibinfo {author} {\bibfnamefont {Maximilian}\
			\bibnamefont {Prilm\"uller}}, \bibinfo {author} {\bibfnamefont {Tobias}\
			\bibnamefont {Huber}}, \bibinfo {author} {\bibfnamefont {Markus}\
			\bibnamefont {M\"uller}}, \bibinfo {author} {\bibfnamefont {Peter}\
			\bibnamefont {Michler}}, \bibinfo {author} {\bibfnamefont {Gregor}\
			\bibnamefont {Weihs}}, \ and\ \bibinfo {author} {\bibfnamefont {Ana}\
			\bibnamefont {Predojevi\ifmmode~\acute{c}\else \'{c}\fi{}}},\ }\bibfield
	{title} {\enquote {\bibinfo {title} {Hyperentanglement of photons emitted by
				a quantum dot},}\ }\href {\doibase 10.1103/PhysRevLett.121.110503} {\bibfield
		{journal} {\bibinfo  {journal} {Phys. Rev. Lett.}\ }\textbf {\bibinfo
			{volume} {121}},\ \bibinfo {pages} {110503} (\bibinfo {year}
		{2018})}\BibitemShut {NoStop}%
	\bibitem [{\citenamefont {Reimer}\ \emph {et~al.}(2019)\citenamefont {Reimer},
		\citenamefont {Sciara}, \citenamefont {Roztocki}, \citenamefont {Islam},
		\citenamefont {Cort{\'e}s}, \citenamefont {Zhang}, \citenamefont {Fischer},
		\citenamefont {Loranger}, \citenamefont {Kashyap}, \citenamefont {Cino} \emph
		{et~al.}}]{reimer19}%
	\BibitemOpen
	\bibfield  {author} {\bibinfo {author} {\bibfnamefont {Christian}\
			\bibnamefont {Reimer}}, \bibinfo {author} {\bibfnamefont {Stefania}\
			\bibnamefont {Sciara}}, \bibinfo {author} {\bibfnamefont {Piotr}\
			\bibnamefont {Roztocki}}, \bibinfo {author} {\bibfnamefont {Mehedi}\
			\bibnamefont {Islam}}, \bibinfo {author} {\bibfnamefont {Luis~Romero}\
			\bibnamefont {Cort{\'e}s}}, \bibinfo {author} {\bibfnamefont {Yanbing}\
			\bibnamefont {Zhang}}, \bibinfo {author} {\bibfnamefont {Bennet}\
			\bibnamefont {Fischer}}, \bibinfo {author} {\bibfnamefont {S{\'e}bastien}\
			\bibnamefont {Loranger}}, \bibinfo {author} {\bibfnamefont {Raman}\
			\bibnamefont {Kashyap}}, \bibinfo {author} {\bibfnamefont {Alfonso}\
			\bibnamefont {Cino}},  \emph {et~al.},\ }\bibfield  {title} {\enquote
		{\bibinfo {title} {High-dimensional one-way quantum processing implemented on
				d-level cluster states},}\ }\href
	{https://www.nature.com/articles/s41567-018-0347-x} {\bibfield  {journal}
		{\bibinfo  {journal} {Nature Physics}\ }\textbf {\bibinfo {volume} {15}},\
		\bibinfo {pages} {148--153} (\bibinfo {year} {2019})}\BibitemShut {NoStop}%
	\bibitem [{\citenamefont {Graham}\ \emph {et~al.}(2015)\citenamefont {Graham},
		\citenamefont {Bernstein}, \citenamefont {Wei}, \citenamefont {Junge},\ and\
		\citenamefont {Kwiat}}]{graham15}%
	\BibitemOpen
	\bibfield  {author} {\bibinfo {author} {\bibfnamefont {Trent~M}\ \bibnamefont
			{Graham}}, \bibinfo {author} {\bibfnamefont {Herbert~J}\ \bibnamefont
			{Bernstein}}, \bibinfo {author} {\bibfnamefont {Tzu-Chieh}\ \bibnamefont
			{Wei}}, \bibinfo {author} {\bibfnamefont {Marius}\ \bibnamefont {Junge}}, \
		and\ \bibinfo {author} {\bibfnamefont {Paul~G}\ \bibnamefont {Kwiat}},\
	}\bibfield  {title} {\enquote {\bibinfo {title} {Superdense teleportation
				using hyperentangled photons},}\ }\href
	{https://www.nature.com/articles/ncomms8185} {\bibfield  {journal} {\bibinfo
			{journal} {Nature communications}\ }\textbf {\bibinfo {volume} {6}},\
		\bibinfo {pages} {1--9} (\bibinfo {year} {2015})}\BibitemShut {NoStop}%
	\bibitem [{\citenamefont {Williams}\ \emph {et~al.}(2017)\citenamefont
		{Williams}, \citenamefont {Sadlier},\ and\ \citenamefont
		{Humble}}]{williams17}%
	\BibitemOpen
	\bibfield  {author} {\bibinfo {author} {\bibfnamefont {Brian~P.}\
			\bibnamefont {Williams}}, \bibinfo {author} {\bibfnamefont {Ronald~J.}\
			\bibnamefont {Sadlier}}, \ and\ \bibinfo {author} {\bibfnamefont {Travis~S.}\
			\bibnamefont {Humble}},\ }\bibfield  {title} {\enquote {\bibinfo {title}
			{Superdense coding over optical fiber links with complete bell-state
				measurements},}\ }\href {\doibase 10.1103/PhysRevLett.118.050501} {\bibfield
		{journal} {\bibinfo  {journal} {Phys. Rev. Lett.}\ }\textbf {\bibinfo
			{volume} {118}},\ \bibinfo {pages} {050501} (\bibinfo {year}
		{2017})}\BibitemShut {NoStop}%
	\bibitem [{\citenamefont {Steinlechner}\ \emph {et~al.}(2017)\citenamefont
		{Steinlechner}, \citenamefont {Ecker}, \citenamefont {Fink}, \citenamefont
		{Liu}, \citenamefont {Bavaresco}, \citenamefont {Huber}, \citenamefont
		{Scheidl},\ and\ \citenamefont {Ursin}}]{steinlechner17}%
	\BibitemOpen
	\bibfield  {author} {\bibinfo {author} {\bibfnamefont {Fabian}\ \bibnamefont
			{Steinlechner}}, \bibinfo {author} {\bibfnamefont {Sebastian}\ \bibnamefont
			{Ecker}}, \bibinfo {author} {\bibfnamefont {Matthias}\ \bibnamefont {Fink}},
		\bibinfo {author} {\bibfnamefont {Bo}~\bibnamefont {Liu}}, \bibinfo {author}
		{\bibfnamefont {Jessica}\ \bibnamefont {Bavaresco}}, \bibinfo {author}
		{\bibfnamefont {Marcus}\ \bibnamefont {Huber}}, \bibinfo {author}
		{\bibfnamefont {Thomas}\ \bibnamefont {Scheidl}}, \ and\ \bibinfo {author}
		{\bibfnamefont {Rupert}\ \bibnamefont {Ursin}},\ }\bibfield  {title}
	{\enquote {\bibinfo {title} {Distribution of high-dimensional entanglement
				via an intra-city free-space link},}\ }\href
	{https://www.nature.com/articles/ncomms15971} {\bibfield  {journal} {\bibinfo
			{journal} {Nature communications}\ }\textbf {\bibinfo {volume} {8}},\
		\bibinfo {pages} {15971} (\bibinfo {year} {2017})}\BibitemShut {NoStop}%
	\bibitem [{\citenamefont {Jin}\ \emph {et~al.}(2019)\citenamefont {Jin},
		\citenamefont {Bourgoin}, \citenamefont {Tannous}, \citenamefont {Agne},
		\citenamefont {Pugh}, \citenamefont {Kuntz}, \citenamefont {Higgins},\ and\
		\citenamefont {Jennewein}}]{Jin:19}%
	\BibitemOpen
	\bibfield  {author} {\bibinfo {author} {\bibfnamefont {Jeongwan}\
			\bibnamefont {Jin}}, \bibinfo {author} {\bibfnamefont {Jean-Philippe}\
			\bibnamefont {Bourgoin}}, \bibinfo {author} {\bibfnamefont {Ramy}\
			\bibnamefont {Tannous}}, \bibinfo {author} {\bibfnamefont {Sascha}\
			\bibnamefont {Agne}}, \bibinfo {author} {\bibfnamefont {Christopher~J.}\
			\bibnamefont {Pugh}}, \bibinfo {author} {\bibfnamefont {Katanya~B.}\
			\bibnamefont {Kuntz}}, \bibinfo {author} {\bibfnamefont {Brendon~L.}\
			\bibnamefont {Higgins}}, \ and\ \bibinfo {author} {\bibfnamefont {Thomas}\
			\bibnamefont {Jennewein}},\ }\bibfield  {title} {\enquote {\bibinfo {title}
			{Genuine time-bin-encoded quantum key distribution over a turbulent
				depolarizing free-space channel},}\ }\href {\doibase 10.1364/OE.27.037214}
	{\bibfield  {journal} {\bibinfo  {journal} {Opt. Express}\ }\textbf {\bibinfo
			{volume} {27}},\ \bibinfo {pages} {37214--37223} (\bibinfo {year}
		{2019})}\BibitemShut {NoStop}%
	\bibitem [{\citenamefont {Marcikic}\ \emph {et~al.}(2004)\citenamefont
		{Marcikic}, \citenamefont {de~Riedmatten}, \citenamefont {Tittel},
		\citenamefont {Zbinden}, \citenamefont {Legr\'e},\ and\ \citenamefont
		{Gisin}}]{marcikic04}%
	\BibitemOpen
	\bibfield  {author} {\bibinfo {author} {\bibfnamefont {I.}~\bibnamefont
			{Marcikic}}, \bibinfo {author} {\bibfnamefont {H.}~\bibnamefont
			{de~Riedmatten}}, \bibinfo {author} {\bibfnamefont {W.}~\bibnamefont
			{Tittel}}, \bibinfo {author} {\bibfnamefont {H.}~\bibnamefont {Zbinden}},
		\bibinfo {author} {\bibfnamefont {M.}~\bibnamefont {Legr\'e}}, \ and\
		\bibinfo {author} {\bibfnamefont {N.}~\bibnamefont {Gisin}},\ }\bibfield
	{title} {\enquote {\bibinfo {title} {Distribution of time-bin entangled
				qubits over 50 km of optical fiber},}\ }\href {\doibase
		10.1103/PhysRevLett.93.180502} {\bibfield  {journal} {\bibinfo  {journal}
			{Phys. Rev. Lett.}\ }\textbf {\bibinfo {volume} {93}},\ \bibinfo {pages}
		{180502} (\bibinfo {year} {2004})}\BibitemShut {NoStop}%
	\bibitem [{\citenamefont {Krenn}\ \emph {et~al.}(2015)\citenamefont {Krenn},
		\citenamefont {Handsteiner}, \citenamefont {Fink}, \citenamefont {Fickler},\
		and\ \citenamefont {Zeilinger}}]{krenn15}%
	\BibitemOpen
	\bibfield  {author} {\bibinfo {author} {\bibfnamefont {Mario}\ \bibnamefont
			{Krenn}}, \bibinfo {author} {\bibfnamefont {Johannes}\ \bibnamefont
			{Handsteiner}}, \bibinfo {author} {\bibfnamefont {Matthias}\ \bibnamefont
			{Fink}}, \bibinfo {author} {\bibfnamefont {Robert}\ \bibnamefont {Fickler}},
		\ and\ \bibinfo {author} {\bibfnamefont {Anton}\ \bibnamefont {Zeilinger}},\
	}\bibfield  {title} {\enquote {\bibinfo {title} {Twisted photon entanglement
				through turbulent air across vienna},}\ }\href {\doibase
		10.1073/pnas.1517574112} {\bibfield  {journal} {\bibinfo  {journal}
			{Proceedings of the National Academy of Sciences}\ }\textbf {\bibinfo
			{volume} {112}},\ \bibinfo {pages} {14197--14201} (\bibinfo {year}
		{2015})}\BibitemShut {NoStop}%
	\bibitem [{\citenamefont {{Da Lio}}\ \emph {et~al.}(2020)\citenamefont {{Da
				Lio}}, \citenamefont {{Bacco}}, \citenamefont {{Cozzolino}}, \citenamefont
		{{Biagi}}, \citenamefont {{Arge}}, \citenamefont {{Larsen}}, \citenamefont
		{{Rottwitt}}, \citenamefont {{Ding}}, \citenamefont {{Zavatta}},\ and\
		\citenamefont {{Oxenløwe}}}]{dalio20}%
	\BibitemOpen
	\bibfield  {author} {\bibinfo {author} {\bibfnamefont {B.}~\bibnamefont {{Da
					Lio}}}, \bibinfo {author} {\bibfnamefont {D.}~\bibnamefont {{Bacco}}},
		\bibinfo {author} {\bibfnamefont {D.}~\bibnamefont {{Cozzolino}}}, \bibinfo
		{author} {\bibfnamefont {N.}~\bibnamefont {{Biagi}}}, \bibinfo {author}
		{\bibfnamefont {T.~N.}\ \bibnamefont {{Arge}}}, \bibinfo {author}
		{\bibfnamefont {E.}~\bibnamefont {{Larsen}}}, \bibinfo {author}
		{\bibfnamefont {K.}~\bibnamefont {{Rottwitt}}}, \bibinfo {author}
		{\bibfnamefont {Y.}~\bibnamefont {{Ding}}}, \bibinfo {author} {\bibfnamefont
			{A.}~\bibnamefont {{Zavatta}}}, \ and\ \bibinfo {author} {\bibfnamefont
			{L.~K.}\ \bibnamefont {{Oxenløwe}}},\ }\bibfield  {title} {\enquote
		{\bibinfo {title} {Stable transmission of high-dimensional quantum states
				over a 2-km multicore fiber},}\ }\href {\doibase 10.1109/JSTQE.2019.2960937}
	{\bibfield  {journal} {\bibinfo  {journal} {IEEE Journal of Selected Topics
				in Quantum Electronics}\ }\textbf {\bibinfo {volume} {26}},\ \bibinfo {pages}
		{1--8} (\bibinfo {year} {2020})}\BibitemShut {NoStop}%
	\bibitem [{\citenamefont {Kim}\ \emph {et~al.}(2006)\citenamefont {Kim},
		\citenamefont {Fiorentino},\ and\ \citenamefont {Wong}}]{kim06}%
	\BibitemOpen
	\bibfield  {author} {\bibinfo {author} {\bibfnamefont {Taehyun}\ \bibnamefont
			{Kim}}, \bibinfo {author} {\bibfnamefont {Marco}\ \bibnamefont {Fiorentino}},
		\ and\ \bibinfo {author} {\bibfnamefont {Franco N.~C.}\ \bibnamefont
			{Wong}},\ }\bibfield  {title} {\enquote {\bibinfo {title} {Phase-stable
				source of polarization-entangled photons using a polarization sagnac
				interferometer},}\ }\href {\doibase 10.1103/PhysRevA.73.012316} {\bibfield
		{journal} {\bibinfo  {journal} {Phys. Rev. A}\ }\textbf {\bibinfo {volume}
			{73}},\ \bibinfo {pages} {012316} (\bibinfo {year} {2006})}\BibitemShut
	{NoStop}%
	\bibitem [{\citenamefont {Fedrizzi}\ \emph {et~al.}(2007)\citenamefont
		{Fedrizzi}, \citenamefont {Herbst}, \citenamefont {Poppe}, \citenamefont
		{Jennewein},\ and\ \citenamefont {Zeilinger}}]{fedrizzi07}%
	\BibitemOpen
	\bibfield  {author} {\bibinfo {author} {\bibfnamefont {Alessandro}\
			\bibnamefont {Fedrizzi}}, \bibinfo {author} {\bibfnamefont {Thomas}\
			\bibnamefont {Herbst}}, \bibinfo {author} {\bibfnamefont {Andreas}\
			\bibnamefont {Poppe}}, \bibinfo {author} {\bibfnamefont {Thomas}\
			\bibnamefont {Jennewein}}, \ and\ \bibinfo {author} {\bibfnamefont {Anton}\
			\bibnamefont {Zeilinger}},\ }\bibfield  {title} {\enquote {\bibinfo {title}
			{A wavelength-tunable fiber-coupled source of narrowband entangled
				photons},}\ }\href {\doibase 10.1364/OE.15.015377} {\bibfield  {journal}
		{\bibinfo  {journal} {Opt. Express}\ }\textbf {\bibinfo {volume} {15}},\
		\bibinfo {pages} {15377--15386} (\bibinfo {year} {2007})}\BibitemShut
	{NoStop}%
	\bibitem [{\citenamefont {Martin}\ \emph {et~al.}(2017)\citenamefont {Martin},
		\citenamefont {Guerreiro}, \citenamefont {Tiranov}, \citenamefont
		{Designolle}, \citenamefont {Fr\"owis}, \citenamefont {Brunner},
		\citenamefont {Huber},\ and\ \citenamefont {Gisin}}]{martin17}%
	\BibitemOpen
	\bibfield  {author} {\bibinfo {author} {\bibfnamefont {Anthony}\ \bibnamefont
			{Martin}}, \bibinfo {author} {\bibfnamefont {Thiago}\ \bibnamefont
			{Guerreiro}}, \bibinfo {author} {\bibfnamefont {Alexey}\ \bibnamefont
			{Tiranov}}, \bibinfo {author} {\bibfnamefont {S\'ebastien}\ \bibnamefont
			{Designolle}}, \bibinfo {author} {\bibfnamefont {Florian}\ \bibnamefont
			{Fr\"owis}}, \bibinfo {author} {\bibfnamefont {Nicolas}\ \bibnamefont
			{Brunner}}, \bibinfo {author} {\bibfnamefont {Marcus}\ \bibnamefont {Huber}},
		\ and\ \bibinfo {author} {\bibfnamefont {Nicolas}\ \bibnamefont {Gisin}},\
	}\bibfield  {title} {\enquote {\bibinfo {title} {Quantifying photonic
				high-dimensional entanglement},}\ }\href {\doibase
		10.1103/PhysRevLett.118.110501} {\bibfield  {journal} {\bibinfo  {journal}
			{Phys. Rev. Lett.}\ }\textbf {\bibinfo {volume} {118}},\ \bibinfo {pages}
		{110501} (\bibinfo {year} {2017})}\BibitemShut {NoStop}%
	\bibitem [{\citenamefont {Franson}(1989)}]{franson89}%
	\BibitemOpen
	\bibfield  {author} {\bibinfo {author} {\bibfnamefont {J.~D.}\ \bibnamefont
			{Franson}},\ }\bibfield  {title} {\enquote {\bibinfo {title} {Bell inequality
				for position and time},}\ }\href {\doibase 10.1103/PhysRevLett.62.2205}
	{\bibfield  {journal} {\bibinfo  {journal} {Phys. Rev. Lett.}\ }\textbf
		{\bibinfo {volume} {62}},\ \bibinfo {pages} {2205--2208} (\bibinfo {year}
		{1989})}\BibitemShut {NoStop}%
	\bibitem [{\citenamefont {Kwiat}\ \emph {et~al.}(1993)\citenamefont {Kwiat},
		\citenamefont {Steinberg},\ and\ \citenamefont {Chiao}}]{kwiat93}%
	\BibitemOpen
	\bibfield  {author} {\bibinfo {author} {\bibfnamefont {P.~G.}\ \bibnamefont
			{Kwiat}}, \bibinfo {author} {\bibfnamefont {A.~M.}\ \bibnamefont
			{Steinberg}}, \ and\ \bibinfo {author} {\bibfnamefont {R.~Y.}\ \bibnamefont
			{Chiao}},\ }\bibfield  {title} {\enquote {\bibinfo {title} {High-visibility
				interference in a bell-inequality experiment for energy and time},}\ }\href
	{\doibase 10.1103/PhysRevA.47.R2472} {\bibfield  {journal} {\bibinfo
			{journal} {Phys. Rev. A}\ }\textbf {\bibinfo {volume} {47}},\ \bibinfo
		{pages} {R2472--R2475} (\bibinfo {year} {1993})}\BibitemShut {NoStop}%
	\bibitem [{\citenamefont {Friis}\ \emph {et~al.}(2019)\citenamefont {Friis},
		\citenamefont {Vitagliano}, \citenamefont {Malik},\ and\ \citenamefont
		{Huber}}]{friis19}%
	\BibitemOpen
	\bibfield  {author} {\bibinfo {author} {\bibfnamefont {Nicolai}\ \bibnamefont
			{Friis}}, \bibinfo {author} {\bibfnamefont {Giuseppe}\ \bibnamefont
			{Vitagliano}}, \bibinfo {author} {\bibfnamefont {Mehul}\ \bibnamefont
			{Malik}}, \ and\ \bibinfo {author} {\bibfnamefont {Marcus}\ \bibnamefont
			{Huber}},\ }\bibfield  {title} {\enquote {\bibinfo {title} {Entanglement
				certification from theory to experiment},}\ }\href
	{https://www.nature.com/articles/s42254-018-0003-5} {\bibfield  {journal}
		{\bibinfo  {journal} {Nature Reviews Physics}\ }\textbf {\bibinfo {volume}
			{1}},\ \bibinfo {pages} {72--87} (\bibinfo {year} {2019})}\BibitemShut
	{NoStop}%
	\bibitem [{\citenamefont {Seri}\ \emph {et~al.}(2017)\citenamefont {Seri},
		\citenamefont {Lenhard}, \citenamefont {Riel\"ander}, \citenamefont
		{G\"undo\ifmmode~\breve{g}\else \u{g}\fi{}an}, \citenamefont {Ledingham},
		\citenamefont {Mazzera},\ and\ \citenamefont {de~Riedmatten}}]{seri17}%
	\BibitemOpen
	\bibfield  {author} {\bibinfo {author} {\bibfnamefont {Alessandro}\
			\bibnamefont {Seri}}, \bibinfo {author} {\bibfnamefont {Andreas}\
			\bibnamefont {Lenhard}}, \bibinfo {author} {\bibfnamefont {Daniel}\
			\bibnamefont {Riel\"ander}}, \bibinfo {author} {\bibfnamefont {Mustafa}\
			\bibnamefont {G\"undo\ifmmode~\breve{g}\else \u{g}\fi{}an}}, \bibinfo
		{author} {\bibfnamefont {Patrick~M.}\ \bibnamefont {Ledingham}}, \bibinfo
		{author} {\bibfnamefont {Margherita}\ \bibnamefont {Mazzera}}, \ and\
		\bibinfo {author} {\bibfnamefont {Hugues}\ \bibnamefont {de~Riedmatten}},\
	}\bibfield  {title} {\enquote {\bibinfo {title} {Quantum correlations between
				single telecom photons and a multimode on-demand solid-state quantum
				memory},}\ }\href {\doibase 10.1103/PhysRevX.7.021028} {\bibfield  {journal}
		{\bibinfo  {journal} {Phys. Rev. X}\ }\textbf {\bibinfo {volume} {7}},\
		\bibinfo {pages} {021028} (\bibinfo {year} {2017})}\BibitemShut {NoStop}%
	\bibitem [{\citenamefont {Pseiner}\ \emph {et~al.}(2020)\citenamefont
		{Pseiner}, \citenamefont {Achatz}, \citenamefont {Bulla}, \citenamefont
		{Bohmann},\ and\ \citenamefont {Ursin}}]{pseiner2020}%
	\BibitemOpen
	\bibfield  {author} {\bibinfo {author} {\bibfnamefont {Johannes}\
			\bibnamefont {Pseiner}}, \bibinfo {author} {\bibfnamefont {Lukas}\
			\bibnamefont {Achatz}}, \bibinfo {author} {\bibfnamefont {Lukas}\
			\bibnamefont {Bulla}}, \bibinfo {author} {\bibfnamefont {Martin}\
			\bibnamefont {Bohmann}}, \ and\ \bibinfo {author} {\bibfnamefont {Rupert}\
			\bibnamefont {Ursin}},\ }\bibfield  {title} {\enquote {\bibinfo {title}
			{Experimental wavelength-multiplexed entanglement-based quantum
				cryptography},}\ }\href {https://arxiv.org/abs/2009.03691} {\bibfield
		{journal} {\bibinfo  {journal} {arXiv preprint arXiv:2009.03691}\ } (\bibinfo
		{year} {2020})}\BibitemShut {NoStop}%
	\bibitem [{\citenamefont {Bennett}\ \emph
		{et~al.}(1996{\natexlab{b}})\citenamefont {Bennett}, \citenamefont
		{DiVincenzo}, \citenamefont {Smolin},\ and\ \citenamefont
		{Wootters}}]{bennett96a}%
	\BibitemOpen
	\bibfield  {author} {\bibinfo {author} {\bibfnamefont {Charles~H.}\
			\bibnamefont {Bennett}}, \bibinfo {author} {\bibfnamefont {David~P.}\
			\bibnamefont {DiVincenzo}}, \bibinfo {author} {\bibfnamefont {John~A.}\
			\bibnamefont {Smolin}}, \ and\ \bibinfo {author} {\bibfnamefont {William~K.}\
			\bibnamefont {Wootters}},\ }\bibfield  {title} {\enquote {\bibinfo {title}
			{Mixed-state entanglement and quantum error correction},}\ }\href {\doibase
		10.1103/PhysRevA.54.3824} {\bibfield  {journal} {\bibinfo  {journal} {Phys.
				Rev. A}\ }\textbf {\bibinfo {volume} {54}},\ \bibinfo {pages} {3824--3851}
		(\bibinfo {year} {1996}{\natexlab{b}})}\BibitemShut {NoStop}%
	\bibitem [{\citenamefont {Treiber}\ \emph {et~al.}(2009)\citenamefont
		{Treiber}, \citenamefont {Poppe}, \citenamefont {Hentschel}, \citenamefont
		{Ferrini}, \citenamefont {Lorünser}, \citenamefont {Querasser},
		\citenamefont {Matyus}, \citenamefont {Hübel},\ and\ \citenamefont
		{Zeilinger}}]{treiber2009}%
	\BibitemOpen
	\bibfield  {author} {\bibinfo {author} {\bibfnamefont {Alexander}\
			\bibnamefont {Treiber}}, \bibinfo {author} {\bibfnamefont {Andreas}\
			\bibnamefont {Poppe}}, \bibinfo {author} {\bibfnamefont {Michael}\
			\bibnamefont {Hentschel}}, \bibinfo {author} {\bibfnamefont {Daniele}\
			\bibnamefont {Ferrini}}, \bibinfo {author} {\bibfnamefont {Thomas}\
			\bibnamefont {Lorünser}}, \bibinfo {author} {\bibfnamefont {Edwin}\
			\bibnamefont {Querasser}}, \bibinfo {author} {\bibfnamefont {Thomas}\
			\bibnamefont {Matyus}}, \bibinfo {author} {\bibfnamefont {Hannes}\
			\bibnamefont {Hübel}}, \ and\ \bibinfo {author} {\bibfnamefont {Anton}\
			\bibnamefont {Zeilinger}},\ }\bibfield  {title} {\enquote {\bibinfo {title}
			{A fully automated entanglement-based quantum cryptography system for telecom
				fiber networks},}\ }\href {\doibase 10.1088/1367-2630/11/4/045013} {\bibfield
		{journal} {\bibinfo  {journal} {New Journal of Physics}\ }\textbf {\bibinfo
			{volume} {11}},\ \bibinfo {pages} {045013} (\bibinfo {year}
		{2009})}\BibitemShut {NoStop}%
	\bibitem [{\citenamefont {Ecker}\ \emph {et~al.}(2019)\citenamefont {Ecker},
		\citenamefont {Bouchard}, \citenamefont {Bulla}, \citenamefont {Brandt},
		\citenamefont {Kohout}, \citenamefont {Steinlechner}, \citenamefont
		{Fickler}, \citenamefont {Malik}, \citenamefont {Guryanova}, \citenamefont
		{Ursin},\ and\ \citenamefont {Huber}}]{ecker19}%
	\BibitemOpen
	\bibfield  {author} {\bibinfo {author} {\bibfnamefont {Sebastian}\
			\bibnamefont {Ecker}}, \bibinfo {author} {\bibfnamefont {Fr\'ed\'eric}\
			\bibnamefont {Bouchard}}, \bibinfo {author} {\bibfnamefont {Lukas}\
			\bibnamefont {Bulla}}, \bibinfo {author} {\bibfnamefont {Florian}\
			\bibnamefont {Brandt}}, \bibinfo {author} {\bibfnamefont {Oskar}\
			\bibnamefont {Kohout}}, \bibinfo {author} {\bibfnamefont {Fabian}\
			\bibnamefont {Steinlechner}}, \bibinfo {author} {\bibfnamefont {Robert}\
			\bibnamefont {Fickler}}, \bibinfo {author} {\bibfnamefont {Mehul}\
			\bibnamefont {Malik}}, \bibinfo {author} {\bibfnamefont {Yelena}\
			\bibnamefont {Guryanova}}, \bibinfo {author} {\bibfnamefont {Rupert}\
			\bibnamefont {Ursin}}, \ and\ \bibinfo {author} {\bibfnamefont {Marcus}\
			\bibnamefont {Huber}},\ }\bibfield  {title} {\enquote {\bibinfo {title}
			{Overcoming noise in entanglement distribution},}\ }\href {\doibase
		10.1103/PhysRevX.9.041042} {\bibfield  {journal} {\bibinfo  {journal} {Phys.
				Rev. X}\ }\textbf {\bibinfo {volume} {9}},\ \bibinfo {pages} {041042}
		(\bibinfo {year} {2019})}\BibitemShut {NoStop}%
	\bibitem [{\citenamefont {Nawaz}\ \emph {et~al.}(2017)\citenamefont {Nawaz},
		\citenamefont {ul~Islam}, \citenamefont {Abbas},\ and\ \citenamefont
		{Ikram}}]{mehwish17}%
	\BibitemOpen
	\bibfield  {author} {\bibinfo {author} {\bibfnamefont {Mehwish}\ \bibnamefont
			{Nawaz}}, \bibinfo {author} {\bibfnamefont {Rameez}\ \bibnamefont
			{ul~Islam}}, \bibinfo {author} {\bibfnamefont {Tasawar}\ \bibnamefont
			{Abbas}}, \ and\ \bibinfo {author} {\bibfnamefont {Manzoor}\ \bibnamefont
			{Ikram}},\ }\bibfield  {title} {\enquote {\bibinfo {title} {Engineering
				quantum hyperentangled states in atomic systems},}\ }\href {\doibase
		10.1088/1361-6455/aa8b40} {\bibfield  {journal} {\bibinfo  {journal} {Journal
				of Physics B: Atomic, Molecular and Optical Physics}\ }\textbf {\bibinfo
			{volume} {50}},\ \bibinfo {pages} {215502} (\bibinfo {year}
		{2017})}\BibitemShut {NoStop}%
	\bibitem [{\citenamefont {Hu}\ and\ \citenamefont {Zhan}(2010)}]{hu10}%
	\BibitemOpen
	\bibfield  {author} {\bibinfo {author} {\bibfnamefont {Bao-Lin}\ \bibnamefont
			{Hu}}\ and\ \bibinfo {author} {\bibfnamefont {You-Bang}\ \bibnamefont
			{Zhan}},\ }\bibfield  {title} {\enquote {\bibinfo {title} {Generation of
				hyperentangled states between remote noninteracting atomic ions},}\ }\href
	{\doibase 10.1103/PhysRevA.82.054301} {\bibfield  {journal} {\bibinfo
			{journal} {Phys. Rev. A}\ }\textbf {\bibinfo {volume} {82}},\ \bibinfo
		{pages} {054301} (\bibinfo {year} {2010})}\BibitemShut {NoStop}%
	\bibitem [{\citenamefont {Brandt}\ \emph {et~al.}(2020)\citenamefont {Brandt},
		\citenamefont {Hiekkam\"{a}ki}, \citenamefont {Bouchard}, \citenamefont
		{Huber},\ and\ \citenamefont {Fickler}}]{brandt:20}%
	\BibitemOpen
	\bibfield  {author} {\bibinfo {author} {\bibfnamefont {Florian}\ \bibnamefont
			{Brandt}}, \bibinfo {author} {\bibfnamefont {Markus}\ \bibnamefont
			{Hiekkam\"{a}ki}}, \bibinfo {author} {\bibfnamefont {Fr\'{e}d\'{e}ric}\
			\bibnamefont {Bouchard}}, \bibinfo {author} {\bibfnamefont {Marcus}\
			\bibnamefont {Huber}}, \ and\ \bibinfo {author} {\bibfnamefont {Robert}\
			\bibnamefont {Fickler}},\ }\bibfield  {title} {\enquote {\bibinfo {title}
			{High-dimensional quantum gates using full-field spatial modes of photons},}\
	}\href {\doibase 10.1364/OPTICA.375875} {\bibfield  {journal} {\bibinfo
			{journal} {Optica}\ }\textbf {\bibinfo {volume} {7}},\ \bibinfo {pages}
		{98--107} (\bibinfo {year} {2020})}\BibitemShut {NoStop}%
	\bibitem [{\citenamefont {Miguel-Ramiro}\ and\ \citenamefont
		{D\"ur}(2018)}]{miguel-ramiro18}%
	\BibitemOpen
	\bibfield  {author} {\bibinfo {author} {\bibfnamefont {J.}~\bibnamefont
			{Miguel-Ramiro}}\ and\ \bibinfo {author} {\bibfnamefont {W.}~\bibnamefont
			{D\"ur}},\ }\bibfield  {title} {\enquote {\bibinfo {title} {Efficient
				entanglement purification protocols for $d$-level systems},}\ }\href
	{\doibase 10.1103/PhysRevA.98.042309} {\bibfield  {journal} {\bibinfo
			{journal} {Phys. Rev. A}\ }\textbf {\bibinfo {volume} {98}},\ \bibinfo
		{pages} {042309} (\bibinfo {year} {2018})}\BibitemShut {NoStop}%
\end{thebibliography}
\end{document}